\documentclass[article,shortnames,nojss]{jss}
\usepackage{color}
\usepackage[dvipsnames,svgnames]{xcolor}
\usepackage{thumbpdf,lmodern}
\usepackage{amsmath,amssymb,amsthm}
\usepackage{bm,bbm}
\usepackage{rotating}
\usepackage[ruled]{algorithm2e}
\usepackage[normalem]{ulem}
\usepackage{siunitx}

\usepackage{color}
\usepackage[dvipsnames,svgnames]{xcolor}

\newcommand{\class}[1]{`\code{#1}'}
\newcommand{\fct}[1]{\code{#1()}}
\renewcommand{\E}{\mathbb{E}}

\newcommand{\abs}[1]{\left\lvert#1\right\rvert}

\theoremstyle{definition}

\newtheorem*{assumption*}{Assumption}

\newcommand{\citepos}[1]{\citeauthor{#1}'s \citeyearpar{#1}}

\author{Stephan Smeekes\\Maastricht University
   \And Ines Wilms\\Maastricht University}
\Plainauthor{Stephan Smeekes, Ines Wilms}

\title{\pkg{bootUR}: An \proglang{R} Package for Bootstrap Unit Root Tests}
\Plaintitle{bootUR: Bootstrap Unit Root Tests in R}
\Shorttitle{\pkg{bootUR}: Bootstrap Unit Root Tests in \proglang{R}}

\Abstract{
Unit root tests form an essential part of any time series analysis. 
We provide practitioners with a single, unified framework for  comprehensive and reliable unit root testing in the \proglang{R} package \pkg{bootUR}. The package's backbone is the popular augmented Dickey-Fuller test paired with a union of rejections principle, which can be performed directly on single time series or multiple (including panel) time series. Accurate inference is ensured through the use of bootstrap methods. The package addresses the needs of both novice users, by providing user-friendly and easy-to-implement functions with sensible default options,
as well as expert users, by giving full user-control to adjust the tests to one's desired settings. Our parallelized \proglang{C++} implementation ensures that all unit root tests are scalable to datasets containing many time series.}

\Plainkeywords{bootstrap, R, time series, unit roots}

\Address{
  Stephan Smeekes, Ines Wilms\\
  Quantitative Economics, School of Business and Economics\\
  Maastricht University\\
  Tongersestraat 53\\
  6211 LM Maastricht, the Netherlands\\
  E-mail: \email{s.smeekes@maastrichtuniversity.nl}, \email{i.wilms@maastrichtuniversity.nl}\\
  URL: \url{https://www.stephansmeekes.nl}, \url{https://sites.google.com/view/iwilms}
}

\begin{document}
\section{Introduction} \label{sec:intro}
In this paper, we introduce the \pkg{bootUR} package \citep{bootUR} for \proglang{R} \citep{R}, which implements several bootstrap tests for unit roots. Unit root testing is an essential part of any statistical analysis of time series. Given the crucial role of unit root testing in time series analysis, surprisingly few \proglang{R} packages exist that allow for easy and comprehensive unit root testing. The \pkg{bootUR} package aims to fill this gap by
offering three major contributions to existing \proglang{R} packages. First, it offers a comprehensive, easy-to-use and reliable set of unit root tests not found as generally in other packages. Second, it offers accurate $p$~values based on bootstrap methods. Third, its functions are not only directly applicable to single time series, but also to datasets consisting of a potentially large set of time series. With these contributions the \pkg{bootUR} package provides practitioners with a single source to fill their unit root testing needs.

Proper handling of unit roots 
is of paramount importance before commencing any form of analysis on the time series of interest. The by far most important use of unit root tests is therefore as a pre-test to determine whether differencing of the series is needed to eliminate the trend and render the time series stationary. 
Ignoring unit roots, or stochastic trends, essentially invalidates any subsequent statistical analysis: the stochastic trend, and associated non-decaying dependence of the present on the far past of a series, yields standard inference
inapplicable. Probably the most famous consequence of ignoring unit roots is the `spurious regression phenomenon', where one finds seemingly important relations (high $R^2$s and highly significant $t$~statistics) between unrelated time series with stochastic trends.
These results have a long history and are well-established and extensively documented in the time series literature. A reader new to unit roots may, for instance, consult \citet{Enders08} for a classical textbook treatment of this spurious regression phenomenon as well as the more general problems associated with unit roots.

Currently, unit root tests are scattered across several packages in the \proglang{R} environment for statistical computing and graphics, making it difficult for a practitioner to find and apply an appropriate and reliable test. 
The most popular unit root test is the classical augmented-Dickey Fuller (ADF) test \citep{DickeyFuller79,dickey1981likelihood}. Implementations of the ADF test are incorporated in various packages, in particular 
\pkg{CADFtest} \citep{lupi2009unit},
\pkg{fUnitRoots} \citep{fUnitRoots}, 
\pkg{tseries} \citep{tseries}, and 
\pkg{urca} \citep{urcabook}.\footnote{The \pkg{mleur}
package \citep{mleurwp} also implements the ADF test, but links to \pkg{urca} for this purpose. The package \pkg{uroot} \citep{uroot} used to have the ADF test implemented but it is no longer supported in the package's current version, hence disregarded from the overview.}

As we will argue in the next section, most `standard' unit root tests, such as the ones implemented in these packages, require practitioners to choose which test to use together with its various model specifications,
that may have a major impact on the performance of the unit root tests. As its first major contribution, the \pkg{bootUR} package instead implements the user-friendly \emph{union of rejections} principle \citep{HLT09, HLT12, SmeekesTaylor12} that relieves the user from the burden of having to choose the right specification and performs this task automatically.

With the exception of the HEGY seasonal unit root test in the \pkg{uroot} package \citep{uroot}, current \proglang{R} implementations of unit root tests rely on asymptotic inference when returning critical values or $p$~values for the unit root test.\footnote{Another exception is the repository \pkg{URT} \citep{URT}, available on GitHub, which includes bootstrap unit root tests. In the remainder, we only focus on packages that are currently maintained on the Comprehensive \proglang{R} Archive Network (CRAN).} 
However, unit root tests are very sensitive to size distortions in smaller samples due to, for example, neglected serial correlation \citep{Schwert89}. Size distortions due to features such as time-varying volatility even persist asymptotically \citep{Cavaliere05}. 
Bootstrap unit root tests have therefore become a commonly used alternative to asymptotic inference. The bootstrap approximates the exact distribution of the unit root test statistic by repeatedly drawing new samples from the original sample, thereby capturing the features of the time series of interest that affect the distribution of the test, which ensures that
bootstrap unit root tests have accurate size properties under very general conditions. The addition of a comprehensive set of bootstrap tests constitutes the second major contribution of our package. 
The `boot' in \pkg{bootUR} stands for bootstrap to emphasize that most
unit roots tests we provide rely on various bootstrap methods for constructing $p$~values.

Finally, most datasets contain multiple, sometimes even many, time series to be tested for unit roots, often leading practitioners to apply unit root tests to each time series separately. Such a practice does not only suffer from multiple testing issues, rejecting several tests by chance alone, but also  disregards similarities between individual time series which, if exploited, could increase the often limited power of the individual tests. 
Although some packages provide joint unit root tests for multivariate or panel data (\pkg{pdR}, \citealp{pdR}; \pkg{plm}, 
\citealp{plmbook, plmJSS}),\footnote{The packages \pkg{PANICr} \citep{PANICr} and \pkg{punitroots} \citep{punitroots} also provide panel unit root tests, but the former has been removed from CRAN and the latter is only available on \proglang{R-Forge} \citep{theussl2009collaborative}.} such tests may increase power but do not allow one to determine the properties of individual series. For this goal one would need tests accounting for multiple testing, but proper implementations of multiple testing corrections are currently lacking for unit root tests. Therefore, the third major contribution of \pkg{bootUR} is to implement easy tools for applying unit root tests to multivariate time series with automatic multiple testing control.

With these contributions, the \pkg{bootUR} package provides a unified framework for easy and comprehensive unit root testing based on the following philosophy.
1) for novice users, the tests should be easy to implement with sensible default options; 
2) those default options should lead to reliable and accurate unit root tests, applicable in general situations; 
3) expert users, familiar with the unit root literature, should be able to easily tweak and adjust the tests to their desired settings; 
4) all tests should be easily scalable to large datasets without additional effort by the user, thereby providing `automatic' functionality.

To accomplish our philosophy, the package has a simple structure, yet it offers users a wide variety of unit root tests. In particular, unit root tests can \textit{directly} be performed on \textit{single} time series or \textit{multiple} time series. To this end, we deliberately created separate functions that serve these purposes: the \fct{adf} function provides the traditional `textbook' ADF test \citep{DickeyFuller79} based on asymptotic $p$~values, while
the functions \fct{boot\_adf} and \fct{boot\_union} can be used for bootstrap-based inference on single time series. 
The function \fct{boot\_ur} is the main function for multiple time series without multiple-testing control, \fct{boot\_sqt} and \fct{boot\_fdr} are set up for multiple time series with multiple-testing control, and the \fct{boot\_panel} function offers a panel unit root test. For each unit root test, the bootstrap method can be chosen by the end-user. To this end, all functions make use of the universal argument \code{bootstrap}. 
Via suitable warning and error messages, user-friendly advise is provided on the (non)-applicability of certain bootstrap methods in certain situations. 
\textit{Model specifications} (such as deterministic components, lag length selection, detrending methods) are under the user's full control, with the option to have them implemented automatically according the union of unit root tests principle which ensures reliable tests across potentially heterogeneous series. Each function contains many options whose syntax is shared across the package, thereby facilitating usability and control by the end-user.

We have also added several functions, based around the core functions above, that aid in the practical implementation of the unit root tests. Most importantly, the function \fct{order\_integration} provides an automatic way to determine the order of integration of each series in a dataset, based on a sequence of one of the aforementioned unit root tests. As it also directly outputs the correctly differenced time series that remove all stochastic trends, it provides the user with the option to conduct the entire unit root pre-analysis with a single command. Additionally, we provide several functions that easily allow the user to assess and visualize properties of the data and outcomes of the tests.

The package is available from the Comprehensive \proglang{R} Archive Network (CRAN) at \url{https://CRAN.R-project.org/package=bootUR}. In addition, the latest (development) version is available on GitHub at \url{https://github.com/smeekes/bootUR}. The core of the package is written in \proglang{C++}, where we make use of the packages \pkg{Rcpp} \citep{eddelbuettel2011rcpp,Rcpp13,Rcpp17} 
and \pkg{RcppArmadillo} \citep{eddelbuettel2014rcpparmadillo} 
to facilitate seamless integration with \proglang{R}. In addition we use the \pkg{RcppParallel} \citep{RcppParallel} package, which provides parallel functionality on all platforms to ensure scalability to large datasets. Version 0.5.0  of the \pkg{bootUR} package and version 4.2.1 of \proglang{R} were used in this paper.

Adhering to the four points of our philosophy not only requires thoughts on how to  implement the tests and design the API, but it also requires a careful choice of the appropriate statistical methods. We therefore first consider the problem from a statistical point of view in Section \ref{sec:URT}, where we discuss the unit root test for single time series and multiple time series, and in Section \ref{sec:boot}, where we discuss the bootstrap methods. We then continue with the package's implementation in Section \ref{sec:package}.
Section \ref{sec:applications} uses three empirical applications to compare \pkg{bootUR}'s unit root functions to implementations in other \proglang{R} packages and illustrate its usefulness for practitioners.
Section \ref{sec:summary} concludes. Appendix \ref{appendix:simstudy} contains a simulation study of the finite sample performance of the methods included in the package.

\section{Unit root tests} \label{sec:URT}
We first discuss unit root tests for individual time series (Section \ref{subsec:iURT}), followed by testing multiple series for unit roots (Section \ref{subsec:mURT}). In our discussion, paralleling \citet{SmeekesWijler20}, we do not focus on theory, but on the issues that arise for practitioners when implementing these tests on their time series.
For a more extensive and theoretical overview of unit root testing, we refer the interested reader to \cite{Choi15}.

\subsection{Unit roots}
Consider the case where we have $T$ observations from a time series $y_t$ ($t=1,\ldots,T$) generated according to the data generating process (DGP) 
\begin{equation}\label{dgp}
\begin{split}
&y_t = x_t + \beta^\top d_t,  \hspace{1cm} x_t = \rho x_{t-1} + u_t, \\
\end{split}
\end{equation}
where $d_t$ are deterministic functions of time. In particular,  three cases are commonly considered: $d_t=0$ (no deterministic components), $d_t=1$ (intercept only), and $d_t=(1,t)^\top$ (intercept and linear trend). The error process $u_t$ is allowed to be serially correlated and heteroskedastic. The presence of serial correlation in $u_t$ has to be accounted for in inference. Typically, $u_t$ is modelled as an invertible infinite order linear process, for instance as
\begin{equation*}
u_t =\sum_{j=0}^\infty \psi_j \epsilon_{t-j} = \sum_{j=1}^{\infty} \phi_j u_{t-j} + \varepsilon_t,
\end{equation*}
where $\varepsilon_t$ is typically assumed to be a martingale difference sequence. This linearity motivates the use of adding lagged differences of the time series to account for the serial dependence, as in the classical ADF test \citep{DickeyFuller79}. However, \citet{paparoditis2018the} show that ADF-type approaches are valid under much more general forms of dependence in $u_t$.

We focus on testing whether or not $y_t$ contains a unit root, that is on testing 
\begin{equation*}
H_0: \rho = 1 \ \text{against} \  \  H_1: \abs{\rho}<1
\end{equation*}
in Equation~\ref{dgp}.
Under the null hypothesis of a unit root, $y_t$ contains a stochastic trend, and equivalently $y_t$ is being said to be integrated of order 1 ($I(1)$), while the alternative postulates that $y_t$ is integrated of order 0 ($I(0)$), which is generally taken as synonymous to $y_t$ being stationary. Here `integrated of order $d$' means that $y_t$ should be differenced $d$ times to achieve a process that does not contain a stochastic trend anymore.\footnote{Although stationary is generally used as synonym for $I(0)$, an $I(0)$ process can still be non-stationary, for instance through a shift in the variance. Despite this distinction, we follow tradition and use `$I(0)$' and `stationary' interchangeably.}

\subsection{Individual unit root tests} \label{subsec:iURT}
To test the null hypothesis of a unit root, the classical ADF test \citep{DickeyFuller79,dickey1981likelihood} remains the pre-dominant choice in practice. For this reason it also forms the backbone of the \pkg{bootUR} package. However, even in its most basic form, practitioners are required to make several non-trivial choices that affect its performance. Table \ref{ADFoverview} summarizes these choices and indicates how the various \proglang{R} packages address each of them. In this section, we first discuss the ADF test and the choices that need to be made, before discussing the union of unit root tests principle proposed by  \cite{HLT09, HLT12} which alleviates many of the concerns.

\begin{table}
\resizebox{0.67\textwidth}{!}{\begin{minipage}{\textwidth}
\begin{tabular}{llcccccccc} \hline \hline
               & Package      & \multicolumn{2}{c}{\pkg{bootUR}}   & \pkg{CADFtest}  & \pkg{fUnitRoots}    & \pkg{tseries}   & \multicolumn{2}{c}{\pkg{urca}}                  \\
               & Function     &\fct{adf} & \fct{boot\_adf} & \fct{CADFtest} &  
               \fct{unitrootTest} & \fct{adf.test} & \fct{ur.df} & \fct{ur.ers} \\
\hline \hline
\textbf{Deterministic}  &  None       & \checkmark  &   \checkmark               &  \checkmark               &       \checkmark                 &       &     \checkmark             &               \\
\textbf{components}  &  Intercept only   &   \checkmark    &   \checkmark              &  \checkmark               &    \checkmark                    &       &   \checkmark                &        \checkmark       \\
\textbf{}  &  Intercept \& linear trend   &  \checkmark    &   \checkmark              & \checkmark                &      \checkmark                  &   \checkmark              &      \checkmark        &        \checkmark       \\
\textbf{}  & User Control & \checkmark   & \checkmark     & \checkmark     & \checkmark           &                 & \checkmark   & \checkmark  \\
\hline
\textbf{Detrending}   &  OLS   &  \checkmark  &    \checkmark           & \checkmark     & \checkmark           & \checkmark      & \checkmark   &               \\
               & QD    &  & \checkmark     &                &                      &                 &              & \checkmark  \\
\hline
\textbf{Lag selection}  & AIC    &  \checkmark      & \checkmark     & \checkmark     &                      &                 & \checkmark   &               \\
               & BIC         & \checkmark  & \checkmark     & \checkmark     &                      &                 & \checkmark   &   \\
               & MAIC        & \checkmark  & \checkmark     & \checkmark     &                      &                 &              &               \\
               & MBIC        &  \checkmark & \checkmark     &                &                      &                 &              &               \\
               & HQC     &     &      &       \checkmark         &                      &                 &              &               \\               
               & Rescaled    & \checkmark  & \checkmark     &                &                      &                 &              &               \\
    & Heuristic &  & & & & \checkmark & & \\           
   & User Control &  \checkmark  & \checkmark     & \checkmark     & \checkmark           & \checkmark      & \checkmark   &    \checkmark        \\               
\hline
\textbf{Inference} & Critical value table  & & & &  \checkmark &  \checkmark & \checkmark& \checkmark\\ 
 & Asymptotic $p$~value  & \checkmark  &                & \checkmark     & \checkmark           & \checkmark      &              &             \\
               & Bootstrap  $p$~value  &  & \checkmark     &                &                      &                 &              &               \\
\hline
\textbf{Other} & Covariate augmentation  & & & \checkmark & & & & \\
\textbf{features} & Explosive alternative & & & & & \checkmark & & \\ \hline 
\end{tabular}
\end{minipage} }
\caption{Overview ADF-test functionalities in existing \proglang{R} packages.} \label{ADFoverview}
\end{table}

\paragraph{ADF test}  Let $\Delta$ be the difference operator defined as $\Delta y_t:= y_{t} - y_{t-1}$. If no deterministic components are present, the ADF regression is given by
\begin{equation}
\Delta y_t = \gamma y_{t-1} + \sum_{j=1}^{p} \phi_{j} \Delta y_{t-j} + \varepsilon_{t}, \hspace{0.5cm} t=p+1, \ldots, T, \label{adf_dc0}
\end{equation}
where the lagged differences of $y_t$ are added to the regression to capture the serial correlation present in $u_t$. Testing the null of a unit root then boils down to testing the significance of the parameter $\gamma$ in Equation~\ref{adf_dc0}.

If the time series $y_t$ is suspected to have deterministic components as well, testing becomes more complicated. The traditional one-step procedure  adds the relevant deterministic components directly in Equation~\ref{adf_dc0}. 
However, the one-step detrending does not lend itself to be extended to other forms of detrending such as the popular generalized least squares (GLS) or quasi-differencing (QD) framework \citep{ERS96} discussed later. Most contemporaneous literature \citep[see e.g.,][]{ChangPark02,HLT09} therefore considers a two-step approach. Here,
a first stage regression is run of $y_t$ on the deterministic components $d_t$, and in the second stage the ADF regression
\begin{equation}
\Delta y_t^d = \gamma y_{t-1}^d + \sum_{j=1}^{p} \phi_{j} \Delta y_{t-j}^d + \varepsilon_{t}^d, \hspace{0.5cm} t=p+1, \ldots, T, \label{adf}
\end{equation}
is run on the residuals of the first stage regression, $y_t^d = y_t - \hat{\beta}^\top d_t$, commonly referred to as the detrended time series.
When estimated by ordinary least squares (OLS), the one-step and two-step detrending are generally used interchangeably and will give very similar results.\footnote{Although they are not equal, and may even have different limit behavior, \cite[cf.][Remark 5]{HLT09}.} 
For the traditional \fct{adf} test, \pkg{bootUR} offers the user the choice to implement one-step as an alternative to the default two-step detrending. This option is mainly provided to allow replicability of results obtained through other software or textbook examples.
The two-step procedure is implemented for all bootstrap unit root tests in the \pkg{bootUR} package, including the \fct{boot\_adf} function for single time series. 

\cite{ERS96} proposed the DF-GLS test, in which 
$\hat{\beta}$ is obtained by a 
GLS regression where the (near-)unit root in $y_t$ is first removed by quasi-differencing; 
the regression is then performed by OLS for $y_t - \left(1 - \frac{c}{T} \right) y_{t-1}$ on $d_t - \left(1 - \frac{c}{T} \right) d_{t-1}$, where $c$ is a parameter that determines how close to differencing the GLS step is; \cite{ERS96} recommend that $c=7$ for the case $d_t = 1$ and $c=13.5$ for the case $d_t = (1,t)^\top$ to yield tests with good power properties. The DF-GLS, or alternatively, DF-QD test,\footnote{To avoid confusion with a `proper' GLS estimation that also takes into account higher-order serial dependence and heteroskedasticty, we follow \cite{HLT09} and refer to this test as the quasi-differenced (QD) test rather than GLS.} 
is often considered to be more powerful than the ADF test, 
although as shown by \citet{MuellerElliott03} and \cite{HLT09} inter alia, the QD test is only more powerful if the initial condition, that is the deviation of the start of the time series from equilibrium, is small. When the initial condition is large, the standard OLS-detrended ADF test is considerably more powerful.
Among other \proglang{R} packages, only \pkg{urca} provides a version of the DF-QD test, see Table \ref{ADFoverview}.

While both options are implemented in the function \fct{boot\_adf} for varying choice of $d_t$, 
the issue of including deterministic components presents the practitioner with two difficult choices: which deterministic components to include, and how to perform the detrending. These choices can have a major impact on the performance.
If too few deterministic components are included, deterministic trends are detected as stochastic trends, and the test becomes inconsistent. On the other hand, adding too many deterministic components reduces the power of the test considerably, and should also be avoided \citep{HLT09}.
Yet, all unit root tests in current \proglang{R} packages ask the user to make these choices. 
Similarly, the initial condition is unobservable, such that the user has to choose
which detrending method to use. 

The \pkg{bootUR} package is, to the best of our knowledge, the first \proglang{R} package which does not force
the user to make these choices, but instead offers the function \fct{boot\_union} for a data-driven alternative via the union of rejections principle introduced by \cite{HLT09, HLT12}. Before discussing this in detail, we first turn to the third  choice a user has to make: selecting the lag length $p$ in Equation~\ref{adf}.

The lag length choice concerns a trade off between size distortions incurred from including too few lags to capture all serial correlation, and power loss incurred from including too many lags. Although theory (and some \proglang{R} packages such as \pkg{tseries}) generally assume $p$ to be a deterministic function of the sample size, in practice data-driven selection allows for a more precise choice of $p$ that can achieve a better
 trade off between size and power.
\pkg{bootUR} offers a set of information criteria  for automatic data-driven lag length selection. 
Information criteria trade off model fit (through the residual sum of squares) and overfitting (through a penalty on the number of parameters). The lag length is estimated as
\begin{equation}
\hat{p}:= \underset{p_{\min} \leq k \leq p_{\max}}{\operatorname{argmin}} IC(k), \hspace{0.5cm} IC(k) = \text{ln}\hat{\sigma}^2_k + k\frac{C_T}{T}, \label{ic}
\end{equation}
where $\hat{\sigma}^2_k:= (T-p_{\max})^{-1}\sum_{t=p_{\max}+1}^{T}(\hat{\varepsilon}_{k,t}^{d})^2$ with $\hat{\varepsilon}_{k,t}^{d}$ the OLS residuals from the ADF regression with lag length $k$ in Equation~\ref{adf}, and $C_T$ is a penalty function that differs according to the information criterion used. We consider two penalties: one corresponding to the Akaike information criterion (AIC; $C_T = 2$) and the other to the Bayesian information criterion (BIC; $C_T = \text{ln}T$).

Next to the original criteria, \pkg{bootUR} also implements their modified variants proposed by \cite{NgPerron01}. These modifications are specifically motivated for lag length selection in the ADF regression. They are given by
\begin{equation}
MIC(k):= \text{ln}\hat{\sigma}^2_k + k\frac{C_T + \xi_T(k)}{T}, \nonumber
\end{equation}
where $\xi_T(k):= (\hat{\sigma}^2_k)^{-1} \hat{\gamma}^2\sum_{t=p_{\max}+1}^{T}(y_{t-1}^d)^2$.
The lag length is then estimated as in Equation~\ref{ic}, with $IC(k)$ replaced by $MIC(k)$.
The modified AIC (MAIC) is obtained by taking $C_T=2$, the modified BIC (MBIC), by taking $C_T=\text{ln}T$. \cite{NgPerron01} show that the \textit{MICs} yield large size improvements over the \textit{ICs} for the purpose of unit root testing. \cite{perron2007simple} recommend to always use the \textit{MICs} with the OLS rather than QD-detrended data (even if the unit root test itself makes use of QD detrending) since this improves the test's power properties; \pkg{bootUR} follows this recommendation. In addition, there are various seemingly minor aspects of how the lag selection is implemented that influence its performance, such as how many observations are used to calculate the residual sum of squares. \citet{NgPerron05} provide a detailed study and guidelines for these choices; \pkg{bootUR} implements the scheme they recommend as optimal.

\cite{CPST15} find that heteroskedasticity affects the performance of information criteria, leading to less accurate choices of $p$ and consequent power loss of the unit root tests. They propose rescaled information criteria, where the time series $y_t$ is rescaled with a nonparametric estimate of its (time-varying) standard deviation, thereby eliminating the heteroskedastictiy. The information criterion is then applied to this rescaled series. These rescaled ICs are generally more powerful in the presence of heteroskedasticity, yet very similar to the original ones without. \pkg{bootUR} therefore performs the rescaling by default (with the option not to consider it) since it is a safe choice and relieves the user of the burden to check whether heteroskedasticity is present.

\paragraph{Union of rejections test}
As mentioned above, choosing the right deterministic components to include and the right detrending method to use, is crucial to obtain tests with good power properties. 
\citet{HLT09, HLT12} take a different approach based on a very simple principle. Roughly speaking, for both specification issues, we have one powerful test and one not powerful test. A logical step would therefore be to perform both tests and reject whenever one of them rejects the null hypothesis - the logic being that the one rejecting is then the powerful one. With two tests performed simultaneously, one must control for multiple testing and adjust the tests with a Bonferroni-type adjustment to control size at the desired level. \citet{HLT09} introduced this union of rejections idea for the two specification issues separately, while \citet{HLT12} combined the two approaches to consider a union of four tests -- intercept only or intercept with trend in combination with OLS or QD detrending -- that guards against both uncertainty over the trend and the initial condition. 

This makes the union test a safe option for quick or automatic unit root testing where careful manual specification is not viable, and makes it therefore very suitable for \pkg{bootUR}'s philosophy that the default option provides a reliable and accurate test, for which no in depth knowledge is needed about either the data or the applicability of various unit root tests. Moreover, it scales easily to large datasets with many series, where careful manual considerations about these specifications are not possible regardless of the expertise of the user.

The \pkg{bootUR} package 
implements the bootstrap version of the union test developed by \citet{SmeekesTaylor12}, which uses the bootstrap both for determining the appropriate size correction and for obtaining the test's $p$~values.
The test statistic takes the form
\begin{equation}\label{eq:URunion}
UR =\min \left(
\frac{s} {c_{QD}^{\mu*} (\alpha)} QD^{\mu}, 
\frac{s} {c_{QD}^{\tau*} (\alpha)} QD^{\tau},
\frac{s} {c_{ADF}^{\mu*} (\alpha)} ADF^{\mu}, 
\frac{s} {c_{ADF}^{\tau*} (\alpha)} ADF^{\tau} \right),
\end{equation}
where $ADF$ and $QD$ are the ADF and QD detrended tests, and superscript $\mu$ and $\tau$ respectively indicating whether the series are demeaned or detrended. The critical values $c_{\cdot}^{\cdot} (\alpha)$ are 
determined in a preliminary bootstrap step as the individual level $\alpha$ critical values of the four tests; weighting with their inverse is needed to bring the four tests on the same scale. The variable $s$ is a scaling factor to which the statistics are scaled. Any $s < 0$ suffices to preserve the left-tail rejection region; in \pkg{bootUR} we scale to -1. This bootstrap union test is made available through the function \fct{boot\_union}. Finally, note that this union-based approach still requires one to select the lag lengths in each of the four ADF regressions. To this end, any of the four information criteria, AIC, BIC, MAIC and MBIC can be used.

Of course, various other unit root tests such as the \cite{phillips1988testing} (PP) test, the KPSS \citep{kwiatkowski1992testing} stationarity test, the HEGY test \citep{hylleberg1990seasonal} for seasonal unit roots, or the covariate-augmented Dickey-Fuller test of \cite{hansen1995rethinking}
exist. While these tests undoubtedly have a lot of value, we intentionally do not implement them in \pkg{bootUR} to avoid overloading the user with choices in tests and options. Instead, we focus on the ADF test which is by far the most popular in practice. Practitioners interested in applying these other unit root tests are recommended to consult the packages \pkg{fUnitRoots}, \pkg{stats}, \pkg{tseries} or \pkg{urca} for the PP test, and \pkg{fUnitRoots}, \pkg{tseries} or \pkg{urca} for the KPSS test. We also want to highlight the \pkg{uroot} package which focuses on the HEGY seasonal unit root test, and the \pkg{CADFtest} package which offers the covariate-augmented ADF test, which exploits correlation with known stationary covariates to improve power. These two packages packages are entirely devoted to these specific unit root tests and already provide excellent functionality for the practitioner in need of these specific tests.

\subsection{Multiple unit root tests} \label{subsec:mURT}
Practitioners often make use of several time series in their analysis, and typically need to test all for unit roots. While performing a unit root test for each series separately is normal practice for a small number of time series, this becomes more complicated if the number of series is large. First, performing many unit root tests simultaneously suffers from   multiple testing issues as the probability of incorrect classifications increases with the number of performed tests. Second, we would like to exploit the similarity between different time series to improve the power of the unit root tests, in particular if the time dimension is relatively small.

In the \pkg{bootUR} package, we consider three different ways to approach the testing problem with multiple time series. First, the simplest option of ignoring the test multiplicity issue by just performing unit root tests separately for each series. To this end, the function \fct{boot\_ur} from \pkg{bootUR} can be used. 
Second, we consider the traditional approach of panel unit root tests, where on pools the information in all series to obtain a more powerful test. The function \fct{boot\_panel} offers such a test. Third, we can consider individual tests but then with appropriate control of multiple testing error rates. \pkg{bootUR} considers two such tests, namely \fct{boot\_sqt} and \fct{boot\_fdr}.

Despite the large literature on this topic, software implementations for multiple unit roots are mostly lacking. While there is some support for panel unit root testing as discussed hereafter, methods to control multiple testing in the context of unit root testing are, to the best of our knowledge, not available. While several general purpose multiple testing packages exist, using these in a proper way with unit root tests requires considerable effort and expertise from the user. For instance, some standard corrections may be overly conservative, such as the Bonferroni correction, or only applicable under specific conditions on the dependence, such as \citepos{BenjaminiHochberg95} method to control the false discovery rate.
As argued by for instance \cite{RSW08}, bootstrap methods for controlling multiple testing allow for general forms of dependence and avoid being too conservative. However, such bootstrap methods need to be integrated with the unit root testing, which is the approach taken in \pkg{bootUR}.

Throughout this section, we use the following notation. Consider $N$ time series for which one would like to test the presence of a unit root. We denote their respective individual unit root test statistics by $UR_i, \ 1\leq i \leq N$. Typically these would correspond to one of the tests discussed in Section \ref{subsec:iURT}. Without loss of generality, we assume that rejections occur for small values of the test statistic.

\paragraph{Panel unit root tests}
Panel unit root tests view the multiple time series as a coherent panel dataset, and exploit the similarity between such time series to pool the information in them and achieve more powerful tests. They have a long tradition in econometrics, see e.g., \cite{breitung2008unit} or \cite{Choi15} for reviews. A typical panel unit root test has the null hypothesis that all series have a unit root. Rejection of this null hypothesis is then typically interpreted as evidence that a `significant proportion' of the series is stationary. 
However, how large that proportion is, or which series are stationary is not revealed by the test. 
\citet{Pesaran12} therefore suggests to use panel unit root tests as an initial screening tool for analyzing multiple series; if the panel unit root test rejects the null, this indicates that the individual series need to examined further; if not, treating the full dataset as $I(1)$ may be a reasonable choice. 

We implement the bootstrap Group-Mean ($GM$) test of \cite{PSU11}
\begin{equation*}
GM = \frac{1}{N}\sum_{i=1}^{N} UR_i,
\end{equation*}
in the function \fct{boot\_panel} which is based on averaging the unit root test statistics $UR_i$ ($1\leq i \leq N$) of the $N$ individual time series. 
This test is valid under very general forms of dependence within the dataset, yet does not require modelling it. 
Panel unit roots tests are scarcely available for \proglang{R} users. Currently, only two packages with panel unit root tests, namely \pkg{plm} and \pkg{pdR}, are being maintained. 
The package \pkg{plm} was the first to offer panel unit root tests and provides the tests introduced in \cite{maddala1999comparative, choi2001unit, levin2002unit, im2003testing}. However, none of them allow for cross-sectional dependence
(see \citealp{kleiber2011panel} for a discussion). 
The package \pkg{pdR} offers the panel unit root test of \cite{chang2002nonlinear} and a panel version of the seasonal unit root test in \cite{hylleberg1990seasonal}.

\paragraph{Multiple testing} 
Multiple testing methods allow practitioners to determine the order of integration for each series in their dataset while controlling the overall probability of making false rejections.
First note that in order to properly rank and compare tests for different series, the individual test statistics should have the same marginal distributions. Then, the ranking
\begin{equation} \label{eq:ranking}
UR_{(1)} \leq \ldots \leq UR_{(R)} \leq UR_{(R+1)} \leq \ldots \leq UR_{(N)},
\end{equation}
corresponds to a ranking from `most significant' to `least significant', when the $i$-th order statistic of $UR_1, \ldots, UR_N$ is denoted by $UR_{(i)}$. 
To ensure the comparability of these statistics, nuisance parameters need to be eliminated from the distribution of the test statistics. \pkg{bootUR} does this automatically for the union test by scaling all test statistics 
in Equation~\ref{eq:URunion} 
towards $s=-1$;
if the user chooses to set specifications manually, it is up to the user to choose them such that any nuisance parameters are eliminated.

The goal is to find an appropriate cut-off point $R$ such that the null of a unit root is rejected for all statistics less than or equal to $UR_{(R)}$, while it is not rejected for all statistics larger. How this threshold is determined, depends on how one controls for multiple testing. \pkg{bootUR} implements two ways to do this:  the sequential testing procedure of \cite{Smeekes15}, which also encompasses the Step-M method of \cite{RomanoWolf05} to control the familywise error rate, and the false discovery rate  (FDR) controlling approach of \citet{RSW08, MoonPerron12}.

\paragraph{Sequential quantile test} \citet{Smeekes15} proposes  the Bootstrap Sequential Quantile Test (BSQT) for multiple unit root testing, that acts as an intermediate between panel unit root testing and full multiple testing control. The method proceeds by sequentially testing groups of time series for unit roots, where the user decides the group sizes. At step 1, we test whether the first $p_1$ series are stationary. Here `first' does not refer to the order in the dataset (which is arbitrary), but to the most significant tests as found via Equation~\ref{eq:ranking}. If the null hypothesis that all $p_1$ units have a unit root cannot be rejected, the test stops. If we do observe a rejection, we move on to the second group where we test if the first $p_2$ are stationary. However, as we already concluded that the first $p_1$ units are stationary, in this second step the actual test is whether the next $p_2 - p_1$ units are stationary as well. We continue this testing procedure until no rejection is observed anymore or we tested all series in the dataset. The BSQT can be performed by using the function \fct{boot\_sqt}.

More formally, let $p_1, \ldots, p_K$ be the number of series to be tested as stationary in each of the steps $k=1,\ldots,K$. In the sequential step $k$ we then test
\begin{equation*}
H_0: p_{k-1} \text{ series are $I(0)$; } \qquad \text{against} \qquad H_1: p_{k} \text{ series are $I(0)$.}
\end{equation*}
As the first test should have as $H_0$ that all units are $I(0)$, $p_0=0$ by default. Furthermore, $p_K=N$ to complete the testing procedure. The number of steps $K$ and the intermediate numbers $p_1, \ldots, p_{K-1}$ can be chosen by the practitioner. Instead of thinking in terms of $p_k$ series, it may be easier to think in terms of quantiles $q_k$, and set $p_k = [q_k N]$. A practitioner may for instance think ``I want to split my series in 10 equally-sized groups.'' In that case the practitioner simply sets $q_k = 0.1 k$.

The choice of $\{p_k\}$ does require input and consideration from the user. 
It can be chosen based on the nature of the dataset and the desired level of precision of the practitioner. \cite{Smeekes15} shows that if $p_k$ units are found to be $I(0)$, the probability that the true number of stationary series lies outside the interval $[p_{k-1}, p_{k+1}]$ is at most the chosen significance level of the test. Finding that $p_k$ series are $I(0)$ should therefore be interpreted as finding that the number of $I(0)$ series is in the interval $[p_{k-1}, p_{k+1}]$. In the end, if $p_2, \ldots, p_{K-1}$ are chosen sensibly and not spaced too far apart, the series that lie in the `uncertain interval' are likely those series which are `just about' significant, and correspond to time series with a $\rho$ parameter very close to 1. The practical consequences of incorrect classification of these series are typically small, as their behavior makes them fit reasonably well in both classes of $I(1)$ and $I(0)$ series.

One special case worth mentioning -- set as the default in \fct{boot\_sqt} -- is when we set $p_k = k$, such that each series gets tested sequentially. 
\cite{Smeekes15} shows that in this case the BSQT method coincides with the popular Step-M method of \cite{RomanoWolf05} to control the familywise error rate (FWE). The FWE is defined as the probability of making at least one false rejection, and is typically controlled via the Bonferroni or \cite{holm1979simple} approach. 

\paragraph{FDR-controlled test}
The false discovery rate (FDR), originally proposed by \citet{BenjaminiHochberg95},  is defined as $FDR = \E \left[\frac{F}{R} \mathbbm{1} (R > 0)\right]$,
where $R$ denote the total number of rejections, and $F$ the number of false rejections. It is more appropriate for larger $N$ than the FWE, as it aims to control the proportion of false rejections to the total, rather than the probability of a single false rejection. \cite{RSW08fdr} develop a bootstrap method to control the FDR, and show that unlike the classical way to control FDR, the bootstrap is appropriate under very general forms of dependence between series. \cite{MoonPerron12} applied this method to unit root testing, and it is their method that is implemented in the \fct{boot\_fdr} function of the \pkg{bootUR} package.
FDR control can be combined with any unit root test specification considered in Section \ref{subsec:iURT}. 

To decide on whether to use BSQT or FDR control, relative sample sizes can be considered. The Monte Carlo comparison of \citet{Smeekes15} reveals that the FDR-controlling test is somewhat more accurate when the sample size $T$ is at least of equal magnitude as the number of time series $N$, whereas the BSQT method is clearly preferable when $T$ is much smaller than $N$, since the FDR-controlling test then suffers from a lack of power.
While \citet{Smeekes15} investigates the performance of (various) bootstrap unit root tests based on the standard ADF test with intercept only, we conduct an additional Monte Carlo simulation experiment in Appendix \ref{appendix:simstudy} that compares the asymptotic \fct{adf} test to the bootstrap tests\fct{boot\_ur}, \fct{boot\_sqt} and \fct{boot\_fdr} based on the union of rejections test, which is the default in the package \pkg{bootUR}.

\section{Bootstrap-based inference} \label{sec:boot}
We mainly rely on bootstrap methods to obtain critical values and/or $p$~values for all of the unit root tests discussed in Section \ref{sec:URT}. In the \pkg{bootUR} package, six bootstrap methods are implemented: the sieve bootstrap (SB), moving block bootstrap (MBB), sieve wild bootstrap (SWB), dependent wild bootstrap (DWB), block wild bootsrap (BWB) and autoregressive wild bootstrap (AWB). Their properties are summarized in Table \ref{boot_prop}, and discussed more extensively below. As immediately apparent from  Table \ref{boot_prop}, any `off-the-shelf' time series bootstrap method may be used to counteract size distortions arising from neglected serial correlation \citep{Schwert89}; whereas a wild bootstrap method is needed to deal with general forms of heteroskedasticity \citep{CavaliereTaylor08, CavaliereTaylor09ER}. General forms of cross-sectional dependence can be captured by any bootstrap method apart from the sieve ones. 

\begin{table}
\begin{tabular}{llcccc} \hline 
Bootstrap method && Serial  & Heteroskedasticity &  Cross-sectional  & Unbalancedness \\
&&  correlation & & dependence  & \\\hline 
SB  && \checkmark & & & \\
MBB && \checkmark & & \checkmark & \\
SWB && \checkmark & \checkmark & & \checkmark \\
DWB && \checkmark & \checkmark & \checkmark & \checkmark\\
BWB && \checkmark & \checkmark & \checkmark & \checkmark\\
AWB && \checkmark & \checkmark & \checkmark & \checkmark\\ \hline
\end{tabular}
\caption{Bootstrap methods and their ability to deal with serial correlation, general forms of heteroskedasticity, cross-sectional dependence and unbalancedness of the data. \label{boot_prop}}
\end{table}

Next to correcting the size of unit root tests, 
bootstrap methods have other advantages. First, the bootstrap offers an automatic $p$~value. 
Second, the bootstrap directly allows for implementation of multiple testing techniques such as those discussed above. 
Third, it guards against misspecification and uncertainty regarding the lag length selection in the ADF. As \pkg{bootUR} re-selects the lag lengths within the bootstrap replications, it automatically takes effects of lag selection into account. This, coupled with the fact that the bootstrap captures any dependence missed by the lagged differences in the ADF regression, adds another layer of protection to the tests.

\subsection{Sieve bootstrap}
The sieve bootstrap (SB) has been extensively considered in the context of unit root testing; see among others \cite{Psaradakis01}, \cite{chang2003sieve}, \cite{PapPol05}, \cite{PSU08} and \cite{smeekes2013detrending}. It estimates the dependence as an autoregressive (AR) process, resamples the residuals of the AR fit, and then re-applies the AR model recursively to place the dependence back into the bootstrap sample. 
\pkg{bootUR} determines the required order of the AR model by the order of the ADF model, combining these in a single step as they should conceptually coincide.

While it is able to capture general forms of serial dependence \citep{KPP11}, it is mostly suited for tests on single time series. \cite{SmeekesUrbain14} show that it is not suited to capture general forms of cross-sectional dependence, making it invalid for joint or multiple testing. 
The \pkg{bootUR} package therefore advises to only use it for unit root testing of a single series or on multivariate series without multiple testing control, throwing a warning to alert the user otherwise. 
When still applied multivariately, 
users should also realize that each time series is required to be observed over the same periods, which we refer to as balanced datasets. This often forces practitioners to delete observations for series that have been observed for a longer period. The reason for this limitation is that resampling step of the sieve bootstrap would reshuffle the missing values, creating bootstrap sample with `holes' in it.

\subsection{Moving block bootstrap}
The moving block bootstrap (MBB) is another traditional bootstrap method that has not only been used for univariate unit root testing in \cite{paparoditis2003residual}, but also for multivariate unit root testing in \cite{MoonPerron12} and \cite{Smeekes15}, as well as for panel unit root testing in \cite{PSU11}. It works by dividing the data in overlapping blocks of data and resampling those blocks to create bootstrap series by laying them end-to-end. The blocks are taken in the time dimension and encompass all series. 
The block length $\ell$ is set automatically by \pkg{bootUR} as a function of the sample size, following a rule proposed by \cite{PSU11} that they showed to perform well in many different circumstances. However, it is easily adjusted by the user to experiment with different lengths and assess the sensitivity of the results for varying block lengths.

The MBB can accommodate any form of serial dependence as long as it `fits' into an adequately sized block, which is a wide class.
Unlike the SB, the MBB can also handle general forms of dependence between series, including but not limited to common factor structures.
From a practical point of view an attractive feature is that it can be applied without requiring one to model the serial and/or cross-sectional dependence.
\cite{PSU11} show its validity for mixed  $I(1)/I(0)$ panel datasets under such general forms of dependence. 
The MBB still has, however, two disadvantages: it cannot handle unbalanced datasets and is sensitive to unconditional heteroskedasticity. 
To handle both issues, users should switch to one of the wild bootstrap methods available in \pkg{bootUR}.

\subsection{Sieve wild bootstrap}
The wild, or multiplier, bootstrap \citep{Mammen93, DavidsonFlachaire08} is known to be robust against general forms of heteroskedasticity, however it cannot handle serial dependence. Nonetheless, if combined with a sieve bootstrap, we get the best of both worlds. That is, by replacing the resampling step applied to the residuals of the AR model with a multiplication by independent and identically distributed (iid) random variables with mean zero and variance one, we obtain the sieve wild bootstrap (SWB). \cite{CavaliereTaylor09ER, cavaliere2009heteroskedastic} and \cite{SmeekesTaylor12} among others apply this sieve wild bootstrap for bootstrap unit root testing. 
The method is perfectly suited to individual unit root testing, but due the AR estimation, suffers from the same inability to capture complex dependence across series as explained by \cite{SmeekesUrbain14} for the SB. 
For the generation of the iid random variables, \pkg{bootUR} follows the unit root papers cited above and uses the normal distribution.

\subsection{Dependent, block and autoregressive wild bootstrap}
The three remaining bootstrap methods implemented in the package are all wild bootstrap methods adjusted to deal with dependence. However unlike the SWB, here the multiplicative random variables themselves are adjusted to be dependent over time. This setup allows these bootstrap methods to capture complex serial and cross-series dependence structures as well as (unconditional) heteroskedasticity. In addition, no resampling takes place for the dependent wild bootstrap, such that missing values `stay in their place' which makes the method applicable to unbalanced datasets. 

The three wild bootstrap methods only differ in how the multiplier variables are made time-dependent. The dependent wild bootstrap method (DWB), originally introduced by \cite{Shao10}, draws random variables from a $T$-dimensional $N(0, \Sigma)$ distribution, where the elements in $\Sigma$ decrease with the distance between them. \cite{Shao10} proposes to use a kernel function to achieve this, along with a bandwidth $\ell$ which ensures that variables more than $\ell$ time points apart are independent. This way $\ell$ has a similar interpretation as the block length in the MBB. \cite{RhoShao19} and \cite{SmeekesUrbain14RM} study the DWB for unit root testing, the latter focusing on multivariate settings.

We consider two more variations. The block wild bootstrap (BWB) \citep{shao2011bootstrap, zhang2014bootstrapping} is a direct alternative to the MBB, where for each block of size $\ell$, we use the same multiplier variable, and the variables are independent between blocks. The autoregressive wild bootstrap (AWB) \citep{SmeekesUrbain14RM, FSU20} generates the multiplier variables as a first-order autoregressive process. Unlike the BWB and DWB who have a block length $\ell$ tuning parameter, the tuning parameter of the AWB is the first-order AR parameter. To be able to use the same tuning parameter $\ell$, we use the conversion formula proposed by \cite{SmeekesUrbain14RM} and \cite{FSU20} that writes the AR parameter as a function of $\ell$, though \pkg{bootUR} also allows to set the AR parameter directly. The default setting for $\ell$ in \pkg{bootUR} uses the same rule as for the MBB, which was also tested for the three wild bootstrap methods by \cite{SmeekesUrbain14RM}. They also provide theoretical results on the validity of these methods under general forms of dependence and heteroskedasticity.

For completeness, in Algorithm \ref{alg:boot} we present the six bootstrap methods and their role in the general bootstrap algorithm. Note that the outcome of the bootstrap algorithm is a collection of bootstrap unit root test statistics $UR_i^b$ for the series $i=1,\ldots, N$ and bootstrap replications $b=1\ldots,B$. How these are then used depends on the multiple testing approach taken. For instance, if we ignore multiple testing, we simply calculate the bootstrap $p$~values
\begin{equation*}
p_i^* = \frac{1}{B} \sum_{b=1}^B I( UR_i^b < UR_i), \qquad i = 1, \ldots, N.
\end{equation*}
For the BSQT and FDR tests more involved processing is needed; for details we refer to \cite{Smeekes15} and \cite{RSW08fdr} respectively.

\begin{algorithm}
\nl Let $y_{i,t}^d = y_{i,t} - d_t \hat{\beta}$, where $\hat{\beta}$ is obtained by OLS\;
\medskip

\nl \For{$i \in \{1,\ldots, N\}$}{
\nl Estimate Equation~\ref{adf} for $\{y_{i,t}\}_{t=1}^T$, determining $p$ by an appropriate criterion, obtaining estimates $(\hat{\gamma}_i, \hat{\phi}_{i,1}, \ldots, \hat{\phi}_{i,p})$\;
\nl Set $\hat{u}_{i,t} = \Delta y_{i,t}^d - \hat{\rho}_i y_{i,t-1}^d$ and $\hat{\varepsilon}_{i,t} = \Delta y_{i,t}^d - \hat{\rho}_i y_{i,t-1}^d - \sum_{j=1}^p \hat{\phi}_{i,j} \Delta y_{i,t-j}^d$, with $y_{-p+1}, \ldots, y_{o} = 0$\;
}
\medskip

\nl \For{$b \in \{1, \ldots, B\}$}{
\medskip

\nl \uIf{SB}{
\nl Generate $s_1, \ldots, s_T$ from a Uniform distribution on $\{1, \ldots, T\}$\;
\nl Set $u_{i,t}^b = \sum_{j=1}^p u_{i,t-j}^b + \varepsilon_{i,t}^b$ with $\varepsilon_{i,t}^b = \hat{\varepsilon}_{i, s_t}$ and $u_{-p+1}, \ldots, u_0 = 0$ for $i=1, \ldots, N$ and $t = 1, \ldots, T$\;

\medskip
} \uElseIf{MBB} {
\nl Generate $s_1^b, \ldots, s_{\lceil T / \ell \rceil}^b$ from a Uniform distribution on $\{1, \ldots, T - \ell + 1\}$\;
\nl \For{$m \in \{1, \ldots, \lceil T / \ell \rceil \}$}{
\nl Set $u_{i,t}^b = \hat{\varepsilon}_{i, s_t^b}$ for $i=1, \ldots, N$ and $t = (m-1)\ell + 1, \ldots, m \ell$\;
}
\medskip
} \uElseIf{SWB}{
\nl Generate $\xi_1^b, \ldots, \xi_{T}^b$ from a $N(0,1)$ distribution\;
\nl Set $u_{i,t}^b = \sum_{j=1}^p u_{i,t-j}^b + \varepsilon_{i,t}^b$ with $\varepsilon_{i,t}^b = \xi_t^b \hat{\varepsilon}_{i, t}$ and $u_{-p+1}, \ldots, u_0 = 0$ for $i=1, \ldots, N$ and $t = 1, \ldots, T$\;
\medskip
} \uElseIf{DWB}{
\nl Generate $\zeta_1^b, \ldots, \zeta_{T}^b$ from a $N(0,1)$ distribution and let $\xi^b = (\xi_1^b, \ldots, \xi_T^b)^\top = \Sigma^{1/2} (\zeta_1^b, \ldots, \zeta_{T}^b)^\top$ with $(\sigma_{s,t})_{s,t =1}^T = K\left(\frac{\abs{s-t}}{\ell}\right)$ for the kernel function $K(\cdot)$ defined in \cite{Shao10}\;
\nl Set $u_{i,t}^b = \xi_t^b \hat{u}_{i, t}$ for $i=1, \ldots, N$ and $t = 1, \ldots, T$\;
\medskip
} \uElseIf{BWB} {
\nl Generate $\xi_1^b, \ldots, \xi_{\lceil T / \ell \rceil}^b$ from a $N(0,1)$ distribution\;
\nl \For{$m \in \{1, \ldots, \lceil T / \ell \rceil \}$}{
\nl Set $u_{i,t}^b = \xi_m^b \hat{u}_{i, t}$ for $i=1, \ldots, N$ and $t = (m-1)\ell + 1, \ldots, m \ell$\;
}
\medskip
} \uElseIf{AWB}{
\nl Generate  $\zeta_2^b, \ldots, \zeta_{T}^b$ from a $N(0,1 - \gamma^2)$ distribution and let $\xi_t^b = \gamma \xi_{t-1}^b + \zeta_t^b$ with $\xi_{1}^b \sim N(0,1)$\;
\nl Set $u_{i,t}^b = \xi_t^b \hat{u}_{i, t}$ for $i=1, \ldots, N$ and $t = 1, \ldots, T$\;
}
\medskip

\nl Set $y_{i,t}^b = \sum_{s=1}^t u_{i,s}^b$\;
\medskip
\nl Let $UR_i^b = UR(y_{i,1}^b, \ldots, y_{i,T}^b)$, where $UR(\cdot)$ denotes the chosen unit root test.
}
\caption{Multivariate Bootstrap Unit Root Tests}\label{alg:boot}
\end{algorithm}

\section[An introduction to the bootUR package]{An introduction to the \pkg{bootUR} package} \label{sec:package}
The \pkg{bootUR} package has a simple structure with 13 user-accessible functions. 
Section \ref{subsec:pkg:datasuitable} presents three functions to check if the data are suitable to be bootstrapped. 
Sections \ref{subsec:pkg:coreindiv} and \ref{subsec:pkg:coremulti} introduce the seven  core functions for unit root testing on respectively individual and multiple time series. 
Section \ref{subsec:pkg:practice} presents three useful functions for determining the order of integration of each series in a particular dataset.

The package's functions will now be presented together with examples of their specific use. 
To this end, we make use of the dataset \code{"MacroTS"}  which contains a collection of 20 macroeconomic time series taken from Eurostat and comes with the package. A complete description of the data can be obtained by typing \code{?MacroTS} in \proglang{R}. The following examples require that the \pkg{bootUR} package and the data have been loaded. As random number generation is required to draw bootstrap samples, we first set the seed of the random number generator to obtain reproducible results:
\begin{CodeChunk}
\begin{CodeInput}
R> library("bootUR")
R> data("MacroTS")
R> set.seed(155776)
\end{CodeInput}
\end{CodeChunk}

\subsection{Checking data suitability} \label{subsec:pkg:datasuitable}
The package \pkg{bootUR} offers three  functions, namely \fct{check\_missing\_insample\_values},  \fct{find\_nonmissing\_subsample} and \fct{plot\_missing\_values},  to 
check if a  particular dataset is suitable to be bootstrapped.
While the bootstrap tests do not work with missing data, unbalanced datasets are generally allowed (see Table \ref{boot_prop}).

The function \fct{check\_missing\_insample\_values} checks if a particular dataset contains missing values. Its usage is extremely simple, as it only requires the data as input, 
\begin{CodeChunk}
\begin{CodeInput}
R> check_missing_insample_values(MacroTS)
\end{CodeInput}
\end{CodeChunk}
which can be a vector, matrix, data frame or in time series format (e.g., \code{ts}, \code{zoo} or \code{xts}). It returns an $N$-dimensional Boolean vector which indicates for each series whether missing values are present (\code{TRUE}) or not (\code{FALSE}).

If a dataset contains series with different starting and end points, the bootstrap methods SWB, DWB, BWB and AWB can still be used. The function \fct{find\_nonmissing\_subsample} lets users check the start and end points of each series as follows:
\begin{CodeChunk}
\begin{CodeInput}
R> sample_check <- find_nonmissing_subsample(MacroTS)
R> sample_check
\end{CodeInput}
\begin{CodeOutput}
$range
      GDP_BE GDP_DE GDP_FR GDP_NL GDP_UK CONS_BE CONS_DE
first      1      1      1      5      1       1       1
last     100    100    100    100    100     100     100
      CONS_FR CONS_NL CONS_UK HICP_BE HICP_DE HICP_FR
first       1       5       1       9       9       9
last      100     100     100     100     100     100
      HICP_NL HICP_UK UR_BE UR_DE UR_FR UR_NL UR_UK
first       9       9     1     1     1     1     1
last      100     100   100   100   100   100   100

$all_equal
[1] FALSE
\end{CodeOutput}
\end{CodeChunk}

The output slot \code{range} returns a $(2\times N)$-matrix displaying the first and last non-missing value for each series, the logical slot \code{all_equal} provides a quick check to see if all time series have the same non-missing indices (\code{TRUE}) or not (\code{FALSE}). 

\begin{figure}
\includegraphics[width=\textwidth]{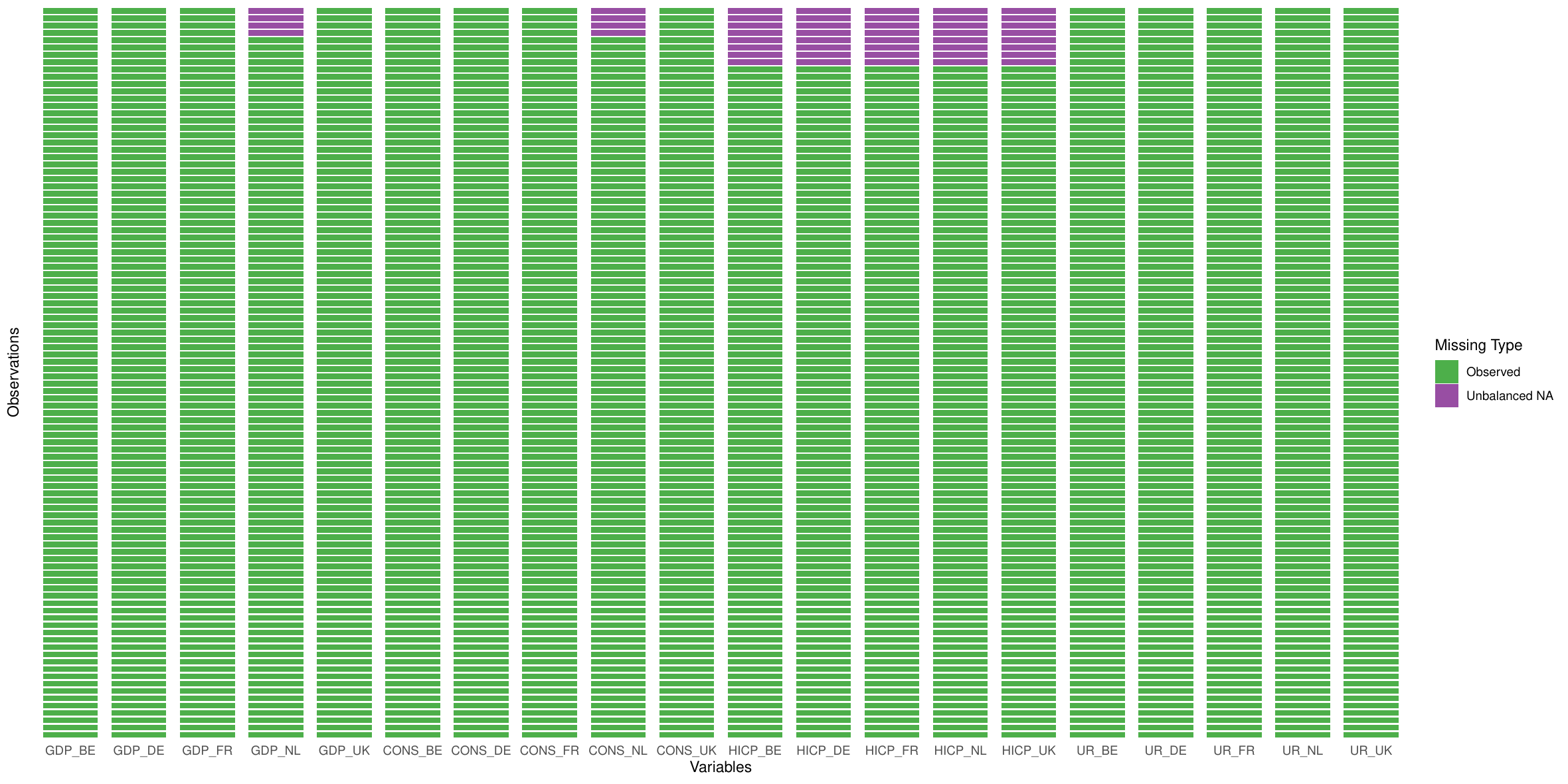}
\caption{\label{MacroTS_missing_plot} Missingness for the dataset \code{MacroTS}.}
\end{figure}

Finally, to display missingness in the dataset, we can use
\begin{CodeChunk}
\begin{CodeInput}
R> plot_missing_values(MacroTS, show_names = TRUE)
\end{CodeInput}
\end{CodeChunk}
which displays present cell values in green, missing values at the start or end (`Unbalanced NAs') in purple and internal missing values in red, see Figure \ref{MacroTS_missing_plot}. Only the latter are problematic for the wild bootstrap methods, while the purple values also need to be avoided for the resampling-based bootstraps.

\subsection{Individual unit root tests} \label{subsec:pkg:coreindiv}
\pkg{bootUR} 
has one function, \fct{adf}, that implements a traditional asymptotic ADF test and
 two functions to perform a bootstrap unit root test on a single series: \fct{boot\_adf} for a standard bootstrap ADF test and \fct{boot\_union} for a bootstrap union test.

We start by discussing 
the standard \fct{adf} function followed by its bootstrap analogue \fct{boot\_adf} with
the many options users can tweak. 
As \pkg{bootUR} shares its syntax across the various functions, the majority of function arguments remains identical across \pkg{bootUR}'s functions, which facilitates usability and control by the end-user. In the remainder, we therefore only highlight the differences compared to the 
\fct{adf} and/or \fct{boot\_adf} function.

\paragraph{Asymptotic ADF test} 
To perform a standard ADF test with asymptotic $p$~value, \pkg{bootUR} offers the \fct{adf} function:
\begin{Code}
adf(data, data_name = NULL, deterministics = "intercept", min_lag = 0, 
 max_lag = NULL, criterion = "MAIC", criterion_scale = TRUE, two_step = TRUE)
\end{Code}
The minimum required input is \code{adf(data)}, where the time series \code{data} can be a vector or a time series object. The optional argument \code{data\_name} allows users to name their data for use in the printed output. 
The remaining set of arguments relates to the ADF regression. The deterministic components can be tweaked via the argument \code{deterministics} with options \code{"none", "intercept", "trend"},  lag length selection in the ADF regression is controlled via the arguments  \code{min\_lag} and \code{max\_lag} that respectively set the minimum and maximum lag length. For the latter, we use the sample size-based default $ \lfloor 12\cdot(T/100)^{1/4} \rfloor$. The information criterion can be selected via the argument \code{criterion} (with options \code{"AIC", "BIC", "MAIC", "MBIC"}) and the argument \code{criterion\_scale} lets practitioners choose to use the rescaled information criteria of \cite{CPST15}. To overwrite data-driven lag selection with a pre-specified lag length, users can simply put both \code{min\_lag} and \code{max\_lag} equal to the desired lag length. Finally, the argument \code{two\_step} allows users to specify whether OLS detrending should be performed following the one-step or two-step procedure, with the latter being the default.

We apply the asymptotic variant of the ADF test with both one-step and two-step OLS detrending to Dutch Gross Domestic Product (GDP). An intercept and linear time trend are added as deterministic components:
\begin{CodeChunk}
\begin{CodeInput}
R> GDP_NL <- MacroTS[, 4]
R> adf_out1 <- adf(data = GDP_NL, deterministics  = "trend", two_step = FALSE)
R> adf_out2 <- adf(data = GDP_NL, deterministics  = "trend", two_step = TRUE)
\end{CodeInput}
\end{CodeChunk}

The function returns a \class{bootUR} class object that inherits from the class \class{htest}. The outcome of the unit root test can be easily read from the console via the \code{print()} function\footnote{We slightly adjust the standard \code{R} output from the \class{htest} class to make it better fit for our unit root tests. Practitioners interested in retrieving the standard \code{R} output from the \class{htest} class, can use \code{stats:::print.htest(adf\_out1)} instead.}: 

\begin{CodeChunk}
\begin{CodeInput}
R> print(adf_out1)
\end{CodeInput}
\begin{CodeOutput}
	One-step ADF test (with trend) on a single time series

data: GDP_NL
null hypothesis: Series has a unit root
alternative hypothesis: Series is stationary

       estimate largest root statistic p-value
GDP_NL                0.9489    -2.487  0.3339
\end{CodeOutput}
\begin{CodeInput}
R> print(adf_out2)
\end{CodeInput}
\begin{CodeOutput}
	Two-step ADF test (with trend) on a single time series

data: GDP_NL
null hypothesis: Series has a unit root
alternative hypothesis: Series is stationary

       estimate largest root statistic p-value
GDP_NL                0.9471    -2.515  0.3202
\end{CodeOutput}
\end{CodeChunk}

The tests differ slightly numerically but both give the same evidence regarding the presence or absence of a unit root. 
Return values such as sample estimate $\hat{\gamma}$, test statistic and $p$~value can be easily retrieved via:
\begin{CodeChunk}
\begin{CodeInput}
R> adf_out1$estimate
\end{CodeInput}
\begin{CodeOutput}
      gamma 
-0.05106074 
\end{CodeOutput}
\begin{CodeInput}
R> adf_out1$statistic
\end{CodeInput}
\begin{CodeOutput}
    tstat 
-2.486711 
\end{CodeOutput}
\begin{CodeInput}
R> adf_out1$p.value
\end{CodeInput}
\begin{CodeOutput}
 p-value 
0.333873 
\end{CodeOutput}
\end{CodeChunk}

Next to the standard slots in \class{htest}, the object  contains 
the additional slots \code{details} and \code{specifications}.
The output slot \code{details} contains an overview of the sample estimate $\hat{\gamma}$ in the slot \code{`individual estimates'},
the test statistic in the slot \code{`individual statistics'}, 
the $p$~value in \code{`individual p-values'}, and the selected number of lagged differences in the ADF regression in the slot \code{`selected lags`}.
The output slot \code{specifications} contains the specifications (deterministics, minimum and maximum lag order, criterion for lag selection, with or without re-scaling and de-trending procedure; one-step or two-step) of the unit root test that is performed.

\paragraph{Bootstrap ADF test} 
To perform a standard ADF bootstrap unit root test on a single series, the \fct{boot\_adf} function can be used.
The function is structured as follows:

\begin{Code}
boot_adf(data, data_name = NULL, bootstrap = "AWB", B = 1999, 
  block_length = NULL, ar_AWB = NULL, deterministics = "intercept", 
  detrend = "OLS", min_lag = 0, max_lag = NULL, criterion = "MAIC", 
  criterion_scale = TRUE, show_progress = TRUE, do_parallel = TRUE, 
  cores = NULL)
\end{Code}

Many arguments used for \fct{adf} re-appear in the \fct{boot\_adf} function and play the same role. We focus the discussion on the new arguments.
The arguments related to the bootstrap specifications include the 
bootstrap method (\code{bootstrap}) and number of bootstrap replications (\code{B}). 
If a user chooses the bootstrap method \code{"MBB"}, \code{"DBB"} \code{"BWB"} or \code{"AWB"}, the desired block length can be controlled via the argument  \code{block\_length}. 
By default, we use \code{block\_length =} $\lfloor1.75 \cdot T^{1/3} \rfloor$, as recommended in \cite{PSU11}.
While for the first three, this argument concerns the genuine block length, for the latter, the block length is transformed into an autoregressive parameter \code{ar\_AWB} via the formula $0.01^{(1/\texttt{l})}$ as in \cite{SmeekesUrbain14RM}; this can be overwritten by setting \code{ar\_AWB} directly. 
The following set of arguments 
relates to the ADF regression and most arguments have a one-to-one correspondence with those in the \fct{adf} function.
For \fct{boot\_adf}, the user can additionally choose the type of detrending via \code{detrend}, with options \code{"OLS"} (default) and \code{"QD"}.

The argument
\code{show\_progress} 
provides live progress updates on the bootstrap. The latter is particularly useful for large values of the argument \code{B}. Finally, the option \code{do\_parallel = TRUE} 
allows the bootstrap to be executed in parallel 
via \pkg{RcppParallel}; the argument \code{cores} allows users to specify how many cores should be used for the parallel loops. By default, all but two cores are used. 

We illustrate the bootstrap ADF test on Dutch GDP, with the sieve bootstrap (\code{bootstrap = "SB"}) as used by \cite{PSU08} and \cite{smeekes2013detrending}. 
We use the same specification for the deterministic components as for the \fct{adf} test above
and detrending is done via OLS.
\begin{CodeChunk}
\begin{CodeInput}
R> boot_adf(data = GDP_NL, bootstrap = "SB",  
+  deterministics  = "trend", detrend = "OLS")
\end{CodeInput}
\begin{CodeOutput}
	SB bootstrap OLS test (with intercept and trend)
	on a single time series

data: GDP_NL
null hypothesis: Series has a unit root
alternative hypothesis: Series is stationary

       estimate largest root statistic p-value
GDP_NL                0.9471    -2.515  0.1311
\end{CodeOutput}
\end{CodeChunk}

As for \fct{adf}, the  function returns an object of class \class{bootUR} and \class{htest}. The bootstrap ADF test returns the same value of the sample estimate and test statistic as for the standard ADF test with two-step OLS detrending as it should, only the $p$~value differs. Also in this case, the test indicates no rejection of the unit root.

\paragraph{Union of rejections test} To perform a bootstrap union unit root test on a single series, the \fct{boot\_union} function can be used.
It shares all its arguments with \fct{boot\_adf} except for \code{deterministics} and \code{detrend} which are omitted since \fct{boot\_union} performs tests for all the possible options,
then combines the outcomes of the four unit root tests as in  Equation~\ref{eq:URunion} to produce a single $p$~value. By combining  these tests, the sample estimate is not available and hence set to  \code{NA} in the output. 
The function introduces one new argument \code{union\_quantile}, which allows the user to choose the $\alpha$ in Equation~\ref{eq:URunion}  used to construct the critical values that serve as weights in the union. If the user has a particular significance level in mind when performing the test, $\alpha$ should be set to this. If not, the default 0.05 is used.

The bootstrap union test for Dutch GDP with the sieve wild bootstrap as proposed by \cite{SmeekesTaylor12} can be obtained via 
\begin{CodeChunk}
\begin{CodeInput}
R> boot_union(data = GDP_NL, bootstrap = "SWB")
\end{CodeInput}
\begin{CodeOutput}
	SWB bootstrap union test on a single time series

data: GDP_NL
null hypothesis: Series has a unit root
alternative hypothesis: Series is stationary

       estimate largest root statistic p-value
GDP_NL                    NA   -0.6701  0.6433
\end{CodeOutput}
\end{CodeChunk}
  
Finally, for the union test, the output slot \code{details} contains an overview of the estimates, test statistics, $p$~values and selected lag orders of the four tests, namely the ADF unit root tests with intercept only or intercept with trend in combination with OLS or GD trending.
 
\subsection{Multiple unit root tests} \label{subsec:pkg:coremulti}
Below, we discuss the various approaches \pkg{bootUR} offers to approach the testing problem with multiple series.

\paragraph{Separate unit root tests} To perform individual ADF tests on multiple time series simultaneously without multiple testing control, the function \fct{boot\_ur} can be used:
\begin{Code}
boot_ur(data, data_name = NULL, bootstrap = "AWB", B = 1999,  
  block_length = NULL, ar_AWB = NULL, level = NULL, union = TRUE, 
  union_quantile = 0.05, deterministics = NULL, detrend = NULL, min_lag = 0,  
  max_lag = NULL, criterion = "MAIC", criterion_scale = TRUE, 
  show_progress = TRUE, do_parallel = TRUE, cores = NULL)
\end{Code}

Compared to the syntax of \fct{boot\_adf}, it has three additional arguments. The argument \code{union} controls whether a bootstrap union test is used (\code{TRUE}) or not (\code{FALSE}). If \code{union = TRUE} (default), the arguments \code{deterministics} and \code{detrend} are ignored, and a warning message is returned if the user would have provided specifications for these anyway. If set to \code{FALSE}, the deterministic components and detrending methods can be specified as for the \fct{boot\_adf} function. The argument \code{union\_quantile} is the same as for \fct{boot\_union}. 
Finally, the optional argument \code{level} offers the user the possibility to give a significance level for the tests to be performed. Doing so adds a slot in the output object which contains a Boolean vector indicating for each series if the unit root null hypothesis is rejected or not. This may be convenient for automated post-processing if the unit root tests are performed as an initial step in an analysis. If a significance level is given, \code{union\_quantile} is overwritten with this value.

Furthermore, since the bootstrap is performed for all series simultaneously, the bootstrap methods \code{"SB"} or \code{"MBB"}, that cannot handle unbalanced datasets, should not be used. If the user were to specify these anyway, the function will revert to splitting the bootstrap up and performing it separately for each time series. A warning message is then returned to alert the user. If a vector (or univariate time series) instead of a matrix (multivariate time series) is given for \code{data}, a single unit root test is performed; in this case the function acts as an alternative to \fct{boot\_adf} and \fct{boot\_union}.\footnote{In fact, internally these functions call \fct{boot\_ur}.}

We illustrate the function's usage by performing individual ADF tests with the \verb|"MBB"| bootstrap on the first five series of the unbalanced dataset \code{MacroTS}, which correspond to the real GDP in Belgium, Germany, France, the Netherlands and the United Kingdom respectively.
\begin{CodeChunk}
\begin{CodeInput}
R> boot_ur_out <- boot_ur(data = MacroTS[, 1:5], bootstrap = "MBB")
\end{CodeInput}

\begin{CodeOutput}
Warning message:
In check_inputs(data = data, boot_sqt_test = boot_sqt_test, 
boot_ur_test = boot_ur_test,  : Missing values cause resampling bootstrap to 
be executed for each time series individually.
\end{CodeOutput}

The warning message alerts the user about the resampling \texttt{"MBB"} bootstrap method being unable to handle unbalanced datasets and the corrective action that is taken to this end.

\begin{CodeInput}
R> boot_ur_out 
\end{CodeInput}
\begin{CodeOutput}
	MBB bootstrap union test on each individual series 
	(no multiple testing correction)

data: MacroTS[, 1:5]
null hypothesis: Series has a unit root
alternative hypothesis: Series is stationary

Tests performed on each series: 
       estimate largest root statistic p-value
GDP_BE                    NA   -0.8135 0.36618
GDP_DE                    NA   -1.1076 0.08804
GDP_FR                    NA   -0.6301 0.76188
GDP_NL                    NA   -0.8211 0.41371
GDP_UK                    NA   -0.7207 0.53877
\end{CodeOutput}
\end{CodeChunk} 
An object of class \class{bootUR} and \class{mult\_htest}, for multiple hypothesis testing, is returned. 
We constructed the class \class{mult\_htest} as an extension of \class{htest} to make it suitable for multiple hypothesis testing. 
In addition to the return values as for the class \class{htest}, it has additional output slots
\code{series.names}, which contains the names of the individual time series, and \code{rejections}, which optionally contains the aforementioned vector with rejections in case this is requested. 

Looking at the results, none of the time series is stationary at the 5\% significance level, as can be seen from the printed $p$~values. 
Information on, for instance, the test statistics and $p$~values can also be accessed directly via:

\begin{CodeChunk}
\begin{CodeInput}
R> boot_ur_out$statistic
\end{CodeInput}
\begin{CodeOutput}
    GDP_BE     GDP_DE     GDP_FR     GDP_NL     GDP_UK 
-0.8135022 -1.1076021 -0.6301366 -0.8210610 -0.7207147 
\end{CodeOutput}
\begin{CodeInput}
R> boot_ur_out$p.value
\end{CodeInput}
\begin{CodeOutput}
    GDP_BE     GDP_DE     GDP_FR     GDP_NL     GDP_UK 
0.36618309 0.08804402 0.76188094 0.41370685 0.53876938 
\end{CodeOutput}
\end{CodeChunk}

\paragraph{Panel unit root test} To perform a panel unit root test, the function \fct{boot\_panel} can be used. It shares its syntax with \fct{boot\_ur}. Unlike for the latter, usage of the \code{"MBB"} or \code{"SB"} bootstrap methods for a panel unit root test on unbalanced datasets will result in an error-- not a warning --since the unbalancedness cannot be reverted. Therefore, users should switch to one of the wild bootstrap methods or balance their dataset. Besides, sieve bootstrap methods can be used, but they are not suited to capture general forms of dependence across units (see Table \ref{boot_prop}). The code therefore warns users against their usage.

We illustrate the usage of the panel unit root test on the five GDP time series with the \verb|"DWB"| bootstrap of \cite{Shao10} and \cite{RhoShao19}:
\begin{CodeChunk}
\begin{CodeInput}
R>  boot_panel(data = MacroTS[, 1:5],  bootstrap = "DWB")
\end{CodeInput}
\begin{CodeOutput}
	Panel DWB bootstrap group-mean union test

data: MacroTS[, 1:5]
null hypothesis: All series have a unit root
alternative hypothesis: Some series are stationary

               estimate largest root statistic p-value
MacroTS[, 1:5]                    NA   -0.8372  0.2986
\end{CodeOutput}
\end{CodeChunk}
Since the null is not rejected at any reasonable level, treating all five GDP series as $I(1)$ is reasonable.

\paragraph{Sequential quantile test}
To perform the BSQT  for multiple unit root testing, the function \fct{boot\_sqt} should be used. It has one additional argument compared to the \fct{boot\_panel} function, namely \code{steps} which sets the group sizes. These can either be set in units or in quantiles. To split the series in, for instance, $K$ equally sized groups, use \code{steps = 0:K / K}. By the convention of \cite{Smeekes15}, the first entry of the vector should be equal to zero, while the second entry indicates the end of the first group, and so on. If the initial zero value or the final value ($N$ or 1 for quantiles)  are accidentally omitted, the function automatically adds them back. The default \code{steps = 0:NCOL(data)} corresponds to the Step-M method of \cite{RomanoWolf05}. Regarding the bootstrap methods, the same warning and error messaging as for the \fct{boot\_panel} apply. To set the significance level of the individual tests performed sequentially, the argument \code{SQT\_level} can be set.

We illustrate the BSQT on the five GDP series with the \code{"AWB"} (default) bootstrap method of \cite{SmeekesUrbain14RM} and \cite{FSU20}: 
\begin{CodeChunk}
\begin{CodeInput}
R> boot_sqt(data = MacroTS[, 1:5]) 
\end{CodeInput}
\begin{CodeOutput}
	AWB bootstrap sequential quantile union test

data: MacroTS[, 1:5]
null hypothesis: Series has a unit root
alternative hypothesis: Series is stationary

Sequence of tests: 
       H0: # I(0) H1: # I(0)  tstat p-value
Step 1          0          1 -1.046  0.3347
\end{CodeOutput}
\end{CodeChunk}
The details on 
the number of series $p_k$ to be tested as stationary in step $k$ under $H_0$ and $H_1$ (first two columns) is printed to the console, together with, the
test-statistic and $p$~value (last two columns) for each of the sequential steps until no rejection occurs. The latter information is also accessible through the output slot \code{details\$SQT}, information on the (non) rejection of the unit root null for each of the series separately can be accessed via the slot \code{rejections}. 

\paragraph{FDR-controlled test}
To perform a multiple unit root test by controlling the FDR, the function \fct{boot\_fdr} should be used. Its arguments are the same as for the other multivariate unit root tests, with the addition of the argument \code{FDR\_level} which sets the FDR level. 
We illustrate it here with the \code{"BWB"} bootstrap method of \cite{shao2011bootstrap} and \cite{SmeekesUrbain14RM}:
\begin{CodeChunk}
\begin{CodeInput}
R> boot_fdr(data = MacroTS[, 1:5], bootstrap = "BWB")
\end{CodeInput}
\begin{CodeOutput}
	BWB bootstrap union test with false discovery rate control

data: MacroTS[, 1:5]
null hypothesis: Series has a unit root
alternative hypothesis: Series is stationary

Sequence of tests: 
         tstat critical value
GDP_DE -0.9814         -1.346
\end{CodeOutput}
\end{CodeChunk}
The procedure developed by \cite{RSW08, MoonPerron12} does not provide $p$~values, therefore critical values are returned instead here. 

\subsection{Determining series' order of integration} \label{subsec:pkg:practice}
Finally, \pkg{bootUR} offers three useful functions for determining the order of integration of each series in  dataset: \fct{order\_integration}, \fct{diff\_mult} and \fct{plot\_order\_integration}.
The main function is \fct{order\_integration} which applies the `Pantula principle' \citep{Pantula89} to determine the order of integration of each series
\begin{Code}
order_integration(data, data_name = NULL, max_order = 2, method = "boot_ur", 
  level = 0.05, plot_orders = FALSE, ...)
\end{Code}
The argument \verb|max_order| sets the maximum order of integration that should be considered for each series. Generally the default of two should generally suffice, with series of order three or higher only very rarely occurring in practice. The user can choose the unit root test through the argument \code{method} depending on whether a single  (\code{"adf"}, \code{"boot\_adf"} or \code{"boot\_union"}) or a multiple time series (\code{"boot\_ur"}, \code{"boot\_sqt"} or \code{"boot\_fdr"}) is considered. All arguments used in these functions can also be passed on via \fct{order\_integration}. 
As setting the order of integration automatically requires ``yes/no'' decisions about the presence of unit roots, a significance level has to be set via the argument \code{level}. 
For \code{"boot\_sqt"} or \code{"boot\_fdr"} this is automatically transformed to \code{SQT\_level} and \code{FDR\_level}, respectively.

The Pantula principle works as follows. It starts by setting $d = \verb|max_order| - 1$ and testing for a unit root on the $\Delta^{d} y_t$ series.
The series for which the unit root null cannot be rejected are classified as $I(d+1)$ and subsequently removed from the dataset. In the next step, $d = d-1$ and the remaining series are tested and classified accordingly. Under the default \verb|max_order = 2|, this second round involves testing the series in levels and classifying them as either $I(1)$ (if the unit root null is not rejected) or $I(0)$ (if the null is rejected). 

The function returns an object of class \class{bootUR} and \class{order\_integration}  with two accessible elements. The slot \code{diff\_data} contains a matrix whose columns are $\Delta^{d_i}y_{i,t}$ with $d_i$ indicating the order of integration of  the $i^{th}$ series  $(i = 1, \ldots, N)$. This matrix is generated by the user-accessible function \code{diff\_mult(data, d)}, where \code{data} is the original dataset and \code{d} is an $N$-dimensional vector indicating each series' order of integration. It contains the same number of rows as the original dataset (since the default setting \code{keep_NAs = TRUE} in \fct{diff\_mult} is used), thereby indicating lost observations as missing. It can be tweaked if a practitioner directly makes use of this function. Note that the object of the input series (e.g., matrix, data frame or time series object) is preserved for the differenced series. The output slot \code{order\_int} makes the vector containing the found orders of integration available to the end-user.

Finally, if the argument \code{plot_orders} in the function \fct{order\_integration} is set to \code{TRUE}, a plot is provided which displays each series' order of integration. To this end, it uses the function \fct{plot\_order\_integration} which takes the object that \fct{order\_integration} outputs as its input. This function is also made accessible if the end-user wishes to further adjust the display of the variable names, legend and colours through its optional arguments \code{show\_names}, \code{show\_legend}, \code{names\_size}, \code{legend\_size} and \code{cols}.

\section{Applications} \label{sec:applications}
We illustrate the methods on three datasets: the income data used in \cite{Smeekes15}, the \verb|MacroTS| dataset which comes with the package, and the FRED-QD dataset, which is widely used for macro-economic analysis.

\paragraph{Replication \cite{Smeekes15}} 
We replicate the empirical application of \cite{Smeekes15} based on income data from the Panel Study of Income Dynamics (PSID) used in \cite{pesaran2007simple}.
The income dataset represents a balanced micro-panel with $T=22$ time periods and $N=181$ units split into three subsamples consisting of college graduates (CLG, $N=58$), high school graduates (HSG, $N=87$), and high school dropouts (HSD, $N=36)$.

We use the functions \fct{boot\_ur}, \fct{boot\_sqt} and \fct{boot\_fdr} with the \texttt{MBB} bootstrap to determine the proportion of stationary units in the PSID dataset with the same specifications as in \cite{Smeekes15}: the bootstrap replications $B$ are set to 4999, the block size  to five, and the lags in the ADF regression are selected using the rescaled MAIC criterion with a maximal lag length of three.

Table \ref{PSID} presents the rejection results of a unit root. \fct{boot\_ur} and \fct{boot\_fdr} indicate that around 30\% of the units are stationary. As discussed in  \cite{Smeekes15}, \fct{boot\_fdr} returns only few units as stationary and thereby likely suffers from low power due to the small sample size. Finally, in line with \cite{Smeekes15}, the proportion of stationary units are fairly similar across subsamples.

\begin{table}
\centering
\begin{tabular}{lccccc} \hline 
&&& \verb|boot_ur| & \verb|boot_sqt| & \verb|boot_fdr|\\ \hline
Total rejections ($N = 181$)    &&&  49 &  60 &   1 \\ 
Proportion of rejections        &&& 0.27 & 0.33 & 0.01 \\ \hline
Rejection in subsamples         &&& & & \\ \hline
  CLG ($N = 58$)                &&&  13 &  16 &   1 \\ 
  Proportion                    &&&  0.22 & 0.28 & 0.02 \\
  HSG ($N = 87$)                &&&  25 &  29 &   0 \\ 
  Proportion                    &&&  0.29 & 0.33 & 0.00 \\
  HSD ($N = 36$)                &&&  11 &  15 &   0 \\ 
  Proportion                    &&& 0.31 & 0.42 & 0 \\
   \hline
\end{tabular}
\caption{Rejections of the unit root null hypothesis in the PSID dataset. \label{PSID}}
\end{table}

\paragraph{MacroTS} The \verb|MacroTS| dataset contains $N=20$ macro-economic time series collected from Eurostat (\href{https://ec.europa.eu/eurostat/data/database}{https://ec.europa.eu/eurostat/data/database}) and is included in the package. Quarterly observations from 1995-2019 ($T=100$) are available on GDP, consumption, inflation and unemployment for Belgium, Germany, France, the Netherlands and the United Kingdom. The dataset is unbalanced, see Figure \ref{MacroTS_missing_plot}.

\begin{figure}
\centering
\includegraphics[width=\textwidth]{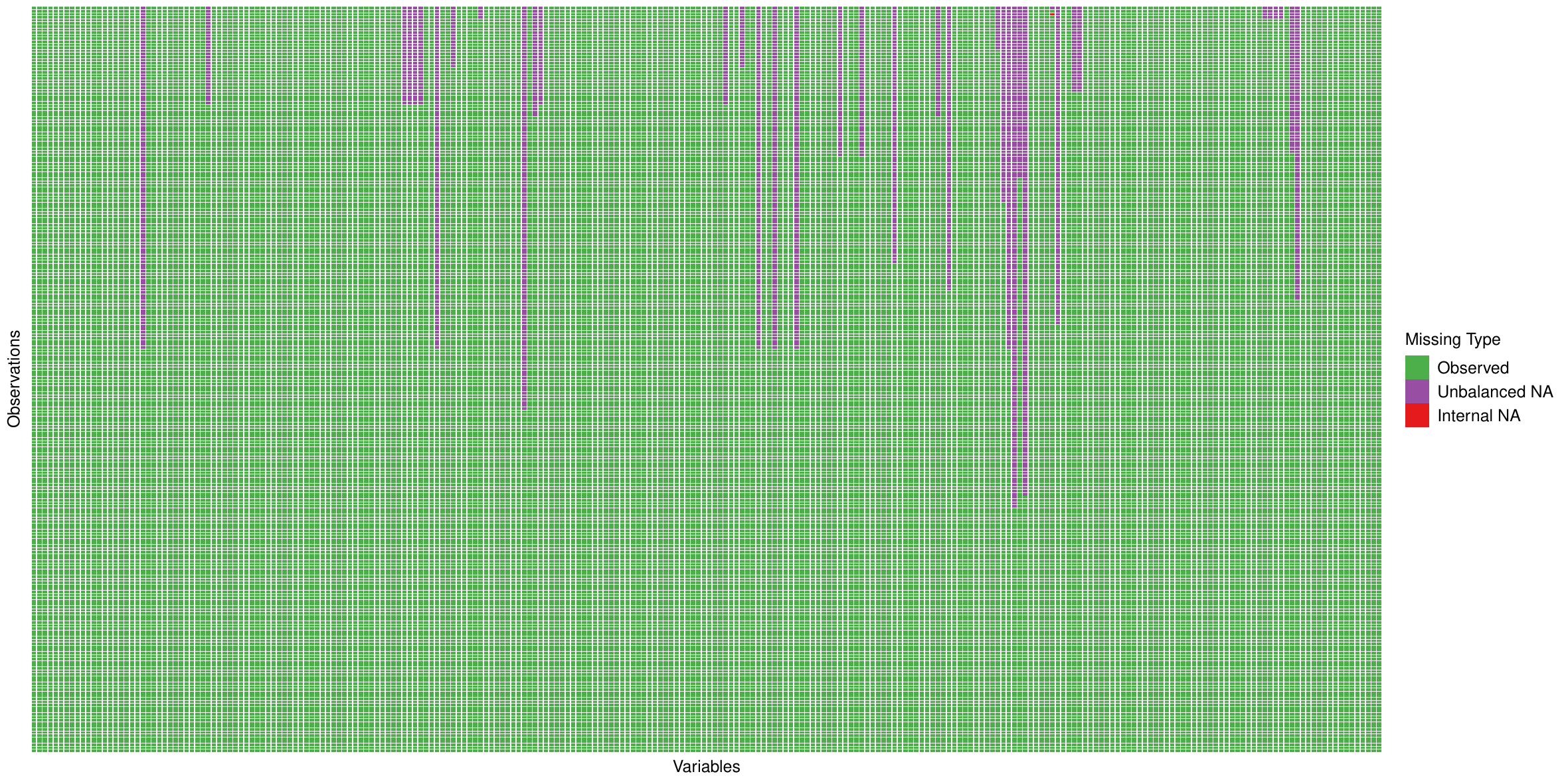}
\caption{\label{FREDQD_NA} Missingness for the dataset FRED-QD.}
\end{figure}

\paragraph{FRED-QD} This is a quarterly version of the monthly Federal Reserve Economic Data  (FRED) database introduced in \cite{McCrackenNg16}. It contains $N=248$ macro-economic time series and was imported into \proglang{R} using the commands
\begin{CodeChunk}
\begin{CodeInput}
R> FRED_url <- url("https://files.stlouisfed.org/files/htdocs/fred-md/
+ quarterly/2020-06.csv")
R> FRED_QD <- read.csv(FRED_url)
\end{CodeInput}
\end{CodeChunk}
This paper uses the data from 1959 Quarter 2 to 2019 Quarter 4 ($T=244$) to avoid possible structural breaks due to the COVID-19 pandemic in 2020. 
If a researcher wishes to import the up-to-date version of the dataset, \code{2020-06.csv} should be changed to \code{current.csv}.
As can be seen from Figure \ref{FREDQD_NA}, the dataset contains one internal NA, since the third observation of variable 188 (\verb|UMCSENTx|: Consumer Expectations) is missing while the second observation is not. \pkg{bootUR} cannot handle internal missing values but this can be easily fixed by setting the second observation to \verb|NA|, which results in the first three observations of this variable being `unbalanced NAs' that can be handled by \pkg{bootUR}.
The resulting dataset then contains 38 macro-economic indicators with missing values at the start of the sample. 
Finally, note that all FRED-QD series have been classified into $I(0), I(1), I(2)$ by  the transformation codes provided in \cite{mccracken2020fred}. However, the authors themselves indicate several discrepancies between these codes and the outcome of unit root tests. We therefore use the transformation codes as a benchmark for the classifications obtained through the unit root tests but do not necessarily consider the classification closest to theirs to be the best.

Since some of the macro-economic series are likely to be $I(2)$, we use the \fct{order\_integration} function (with its defaults) to implement the Pantula principle. All unit root tests in the \pkg{bootUR} are performed with their default settings, which means that union tests are performed with the \verb|AWB| bootstrap method, and lag length selection is done via the re-scaled MAIC. Throughout this section, a significance level of 5\% is used.
For \fct{boot\_sqt}, the default (i.e., Step-M method) is reported as well as results for evenly spaced 0.1 quantiles (\code{steps = 0:10/10}, for \code{MacroTS}), and 0.05 quantiles (\code{steps = 0:20/20}, for FRED-QD).

We compare \pkg{bootUR}'s unit root tests to the \proglang{R} packages reported in Table \ref{ADFoverview}. We hereby use the following specifications:
For the function \fct{CADFtest} (package \pkg{CADFtest}), we perform ADF regressions with intercept and trend (\verb|type = "trend"|), and lag length selection with  MAIC (\verb|criterion = "MAIC"|) thereby considering a maximum of $\lfloor 12\cdot(T/100)^{1/4} \rfloor$ lags, set via the argument \verb|max.lag.y|. These lag length specifications correspond to the defaults used in \pkg{bootUR}.  
For \fct{unitrootTest} (package \pkg{fUnitRoots}), we perform ADF regressions with intercept and trend (\verb|type = "ct"|). By default, one lagged difference is included.
For  \fct{adf.test} (package \pkg{tseries}), we use its default settings which implies ADF-regressions with intercept and trend and the number of lags fixed to $\lfloor (T-1)^{1/3}\rfloor$, a deterministic function of the sample size.
For \fct{ur.df} (package \pkg{urca}), we use ADF-regressions with intercept and trend (\verb|type="trend"|), lag length selection via  AIC (\verb|selectlags = "AIC"|), thereby considering a maximum of  $\lfloor 12\cdot(T/100)^{1/4} \rfloor$ lags, set via the argument \verb|lags|.
Finally, for \fct{ur.ers} (package \pkg{urca}), we use an intercept and trend for detrending (\verb|model| = "trend"). By default, four lagged differences are included in the ADF-regression.
Unlike the other packages,  \pkg{urca} only comes with critical values to judge the significance of the unit root test, the $p$~value is not reported, see Table \ref{ADFoverview}. As discussed in \cite{lupi2009unit}, the $p$~value reported under \fct{summary} is computed using the $t$~distribution, which is incorrect under the unit root null.
Finally, only the packages \pkg{CADFtest} and \pkg{fUnitRoots} can handle missing values, for the other packages, we removed missing values prior to performing the unit root tests.

Before applying the various unit root tests to the two datasets, we perform the \fct{boot\_panel} (with default settings) to all series taken in first differences, and to all series in levels. Table \ref{panel_test_applications} reports the $p$~values of the panel unit root tests. 
For  both datasets, the panel unit root tests on the series in first differences indicates that the unit root null is rejected, thereby indicating that a `significant proportion' of the series is stationary in first differences (hence not $I(2)$). 
The panel unit root test on the series in levels indicates non-rejection of the unit root null. 

\begin{table}
\centering
\begin{tabular}{lcccc} \hline 
Series                  && \verb|MacroTS| && FRED-QD \\ \hline 
in first differences    &&      $<0.001$  &&   $<0.001$      \\  
in levels               &&      0.097    &&    0.197       \\ \hline 
\end{tabular}
\caption{$p$~values of the panel unit root test on all series in differences and all series in levels, for both the \code{MacroTS} and FRED-QD dataset. \label{panel_test_applications}}
\end{table}

\begin{figure}
\includegraphics[width=\textwidth]{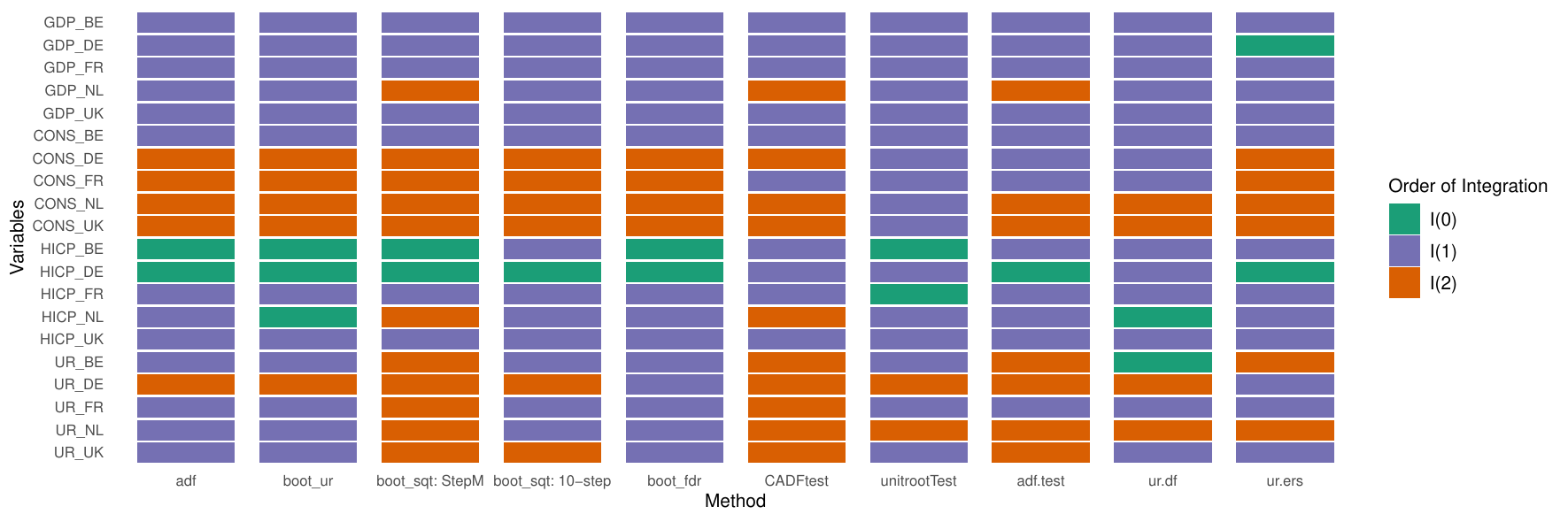}
\caption{\label{MacroTS_order_intergration} Classification of the \code{MacroTS} dataset into $I(0), I(1), I(2)$.}
\end{figure}

\begin{figure}
\centering
\includegraphics[width=\textwidth]{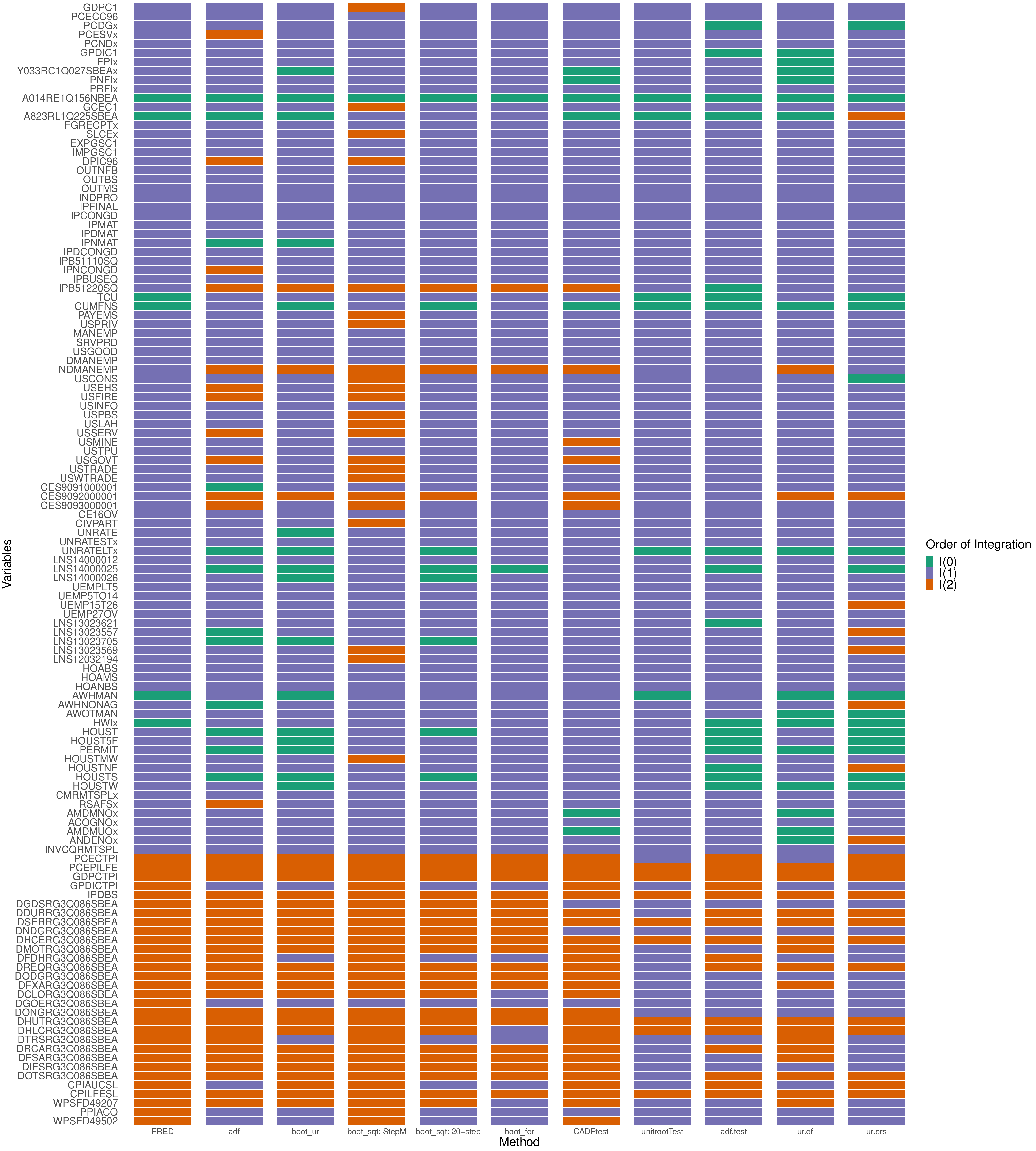}
\caption{\label{FRED_QD_order_intergration_1to124} Classification of the first half of time series in the FRED-QD dataset into $I(0), I(1), I(2)$.}
\end{figure}

\begin{figure}
\centering
\includegraphics[width=\textwidth]{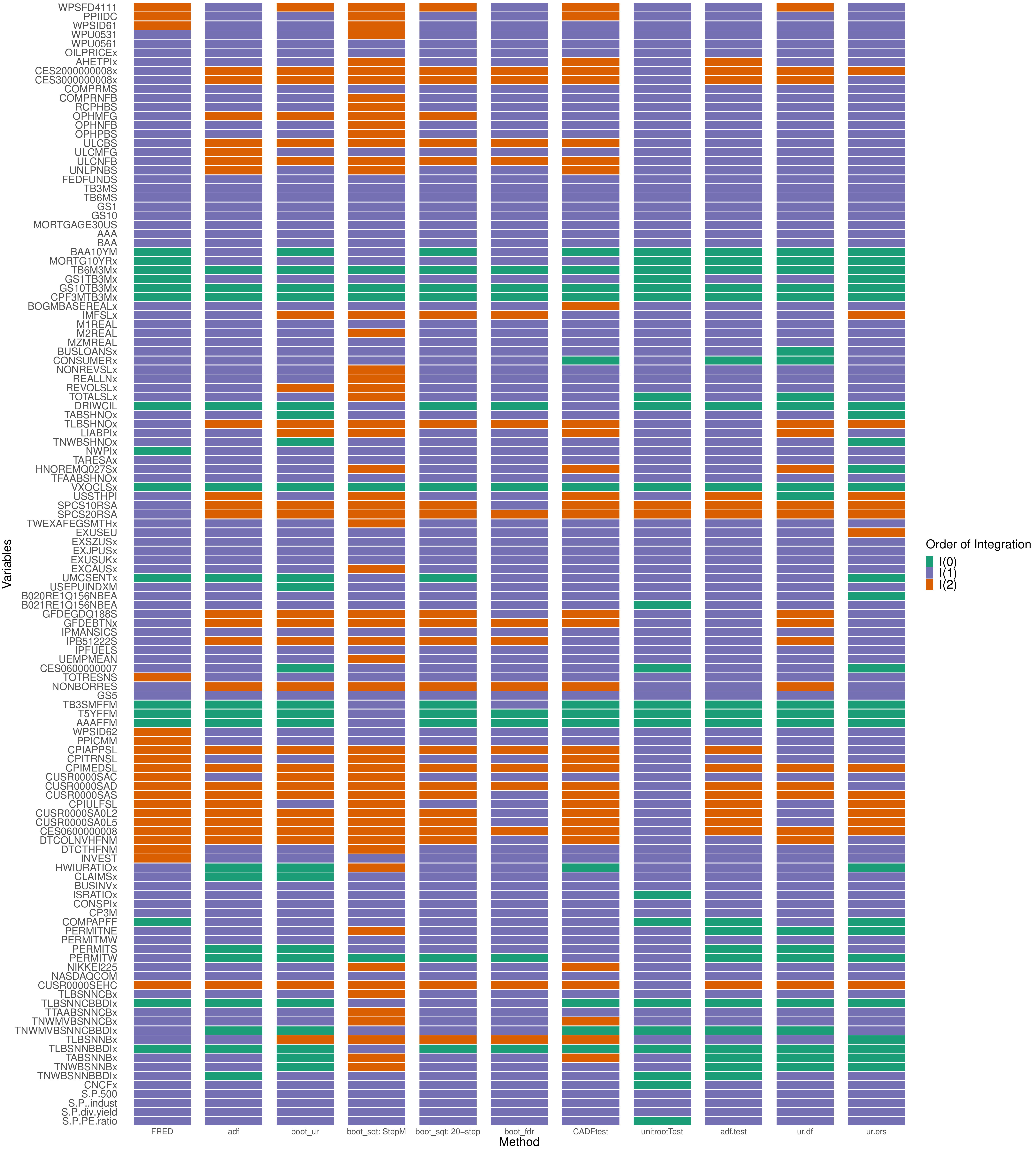}
\caption{\label{FRED_QD_order_intergration_125to248} Classification of the second half of time series in the FRED-QD dataset into $I(0), I(1), I(2)$.}
\end{figure}

To shed further light on the order of integration for each of the individual series, the bootstrap unit root tests are applied and compared to the implementations from other \proglang{R} packages. Figure \ref{MacroTS_order_intergration} presents the obtained orders of integration on the \verb|MacroTS| dataset, Figures \ref{FRED_QD_order_intergration_1to124} and \ref{FRED_QD_order_intergration_125to248} on the FRED-QD dataset. Note that on such a large dataset, performing the bootstrap tests -- even twice in \fct{order\_integration} if \code{d=2} -- can become computationally intensive. Table \ref{tab:runningtime} gives the computation times relative to our \fct{adf} function.\footnote{As recorded on a MacBook Pro (macOS 11.6.7) with Quad-Core Intel Core i5 2,4 GHz processor.} While the \fct{adf} function took just over a second to run, it took roughly between 7 and 8 minutes to perform each of the bootstrap tests. There is therefore a price to pay for bootstrap accuracy. A simple way to save computation time, while still preserving reasonable accuracy, is to reduce the number of bootstrap replications, by altering the argument \code{B}.

\begin{table}[h]
\centering
\caption{Computing time of the unit root tests relative to the benchmark \fct{adf} test for the FRED-QD dataset on a MacBook Pro (macOS 11.6.7) with Quad-Core Intel Core i5 2,4 GHz processor under the default parallel computing settings.} \label{tab:runningtime}
\begin{tabular}{lS[table-format=3.2]lS[table-format=3.2]} \hline 
\multicolumn{2}{c}{Non-bootstrap tests} & \multicolumn{2}{c}{Bootstrap tests}\\ \hline 
\code{CADFtest}         & 134.27 & \code{boot\_ur} & 343.75 \\
\code{unitrootTest}     & 1.70 &   \code{boot\_sqt} & 306.68 \\
\code{adf.test}         & 0.50 &   \code{boot\_sqt\_20step} & 343.01\\
\code{ur.df}            & 5.71 &   \code{boot\_fdr} & 351.72\\
\code{ur.ers}           & 0.90 &  &  \\ \hline 
Benchmark: \code{adf} & 1.33 & (absolute time in seconds)\\
\hline
\end{tabular}
\end{table}
Globally speaking, most unit root tests agree upon a series' classification into $I(0), I(1), I(2)$, which is comforting. Still, several interesting remarks can be made.
First, the results of \fct{boot\_ur} are fairly similar to \fct{adf}, \fct{boot\_sqt} and \fct{boot\_fdr} but it classifies a considerable amount of  series as $I(0)$ instead of $I(1)$ on the FRED-QD dataset. This illustrates that ignoring multiple testing can quickly lead to a considerable number of misclassifications on such large datasets. Second, among the \fct{boot\_sqt} procedures, the default Step-M method tends to classify more series as $I(2)$ than the other procedure. 
On the smaller \verb|MacroTS| dataset the two versions of the \fct{boot\_sqt} show more agreement than on the larger FRED-QD dataset. Third, \fct{boot\_fdr} tends to classify more series as $I(1)$ than the other tests. For a more elaborate discussion of this tendency, we refer the interested reader to \cite{SmeekesWijler20}. Fourth, among the unit root tests from the other \proglang{R} packages, \fct{CADFtest} produces most similar results to \pkg{bootUR}. The function \fct{unitrootTest} detects far less series as $I(2)$. 

\section{Summary} \label{sec:summary}
This paper presents the \proglang{R} package \pkg{bootUR} that provides a unified framework for bootstrap unit root testing on single and multiple time series. To this end, the package builds upon the popular ADF test with a union of rejections principle.
Unlike existing packages on unit root tests, \pkg{bootUR}
(i) provides a large collection of easy-to-use, fully-controllable and reliable unit root tests, including the union of rejections test which is set as default to enable quick, automatic unit root testing, 
(ii) ensures accurate inference through bootstrap methods with easy-to-read output (including $p$~values),
(iii) allows for testing the presence of unit roots in datasets containing many time series by relying on fast \proglang{C++} implementations.

\section*{Acknowledgments}
We thank the editors and  referees for their thorough review and highly appreciate their constructive comments which substantially improved the quality of the package and paper.
The first author was financially supported by the Netherlands Organization for Scientific Research (NWO) under grant number 452-17-010, the second author by  the European Union's Horizon 2020 research and innovation programme under the Marie Sk\l{}odowska-Curie grant agreement  No 832671. We gratefully acknowledge the comments and checks provided by Robert Adamek, Rui Jorge Almeida, Nalan Ba\c{s}t\"urk, Caterina Schiavoni and Eti\"enne Wijler on earlier versions of the package. All remaining errors are our own.

\bibliography{refs}

\begin{thebibliography}{81}
\newcommand{\enquote}[1]{``#1''}
\providecommand{\natexlab}[1]{#1}
\providecommand{\url}[1]{\texttt{#1}}
\providecommand{\urlprefix}{URL }
\expandafter\ifx\csname urlstyle\endcsname\relax
  \providecommand{\doi}[1]{doi:\discretionary{}{}{}#1}\else
  \providecommand{\doi}{doi:\discretionary{}{}{}\begingroup
  \urlstyle{rm}\Url}\fi
\providecommand{\eprint}[2][]{\url{#2}}

\bibitem[{Allaire \emph{et~al.}(2021)Allaire, Francois, Ushey, Vandenbrouck,
  Geelnard, and {Intel}}]{RcppParallel}
Allaire J, Francois R, Ushey K, Vandenbrouck G, Geelnard M, {Intel} (2021).
\newblock \emph{\pkg{RcppParallel}: Parallel Programming Tools for
  `\pkg{Rcpp}'}.
\newblock \proglang{R} package version 5.1.4,
  \urlprefix\url{https://CRAN.R-project.org/package=RcppParallel}.

\bibitem[{Benjamini and Hochberg(1995)}]{BenjaminiHochberg95}
Benjamini Y, Hochberg Y (1995).
\newblock \enquote{Controlling the False Discovery Rate: A Practical and
  Powerful Approach to Multiple Testing.}
\newblock \emph{Journal of the Royal Statistical Society B}, \textbf{57}(1),
  289--300.

\bibitem[{Breitung and Pesaran(2008)}]{breitung2008unit}
Breitung J, Pesaran MH (2008).
\newblock \enquote{Unit Roots and Cointegration in Panels.}
\newblock In \emph{The Econometrics of Panel Data}, pp. 279--322.
  Springer-Verlag.

\bibitem[{Bronder(2016)}]{PANICr}
Bronder S (2016).
\newblock \emph{\pkg{PANICr}: PANIC Tests of Nonstationarity}.
\newblock \proglang{R} package version 1.0.0,
  \urlprefix\url{https://CRAN.R-project.org/package=PANICr}.

\bibitem[{Cavaliere(2005)}]{Cavaliere05}
Cavaliere G (2005).
\newblock \enquote{Unit Root Tests Under Time-Varying Variances.}
\newblock \emph{Econometric Reviews}, \textbf{23}(3), 259--292.

\bibitem[{Cavaliere \emph{et~al.}(2015)Cavaliere, Phillips, Smeekes, and
  Taylor}]{CPST15}
Cavaliere G, Phillips PCB, Smeekes S, Taylor AMR (2015).
\newblock \enquote{Lag Length Selection for Unit Root Tests in the Presence of
  Nonstationary Volatility.}
\newblock \emph{Econometric Reviews}, \textbf{34}(4), 512--536.

\bibitem[{Cavaliere and Taylor(2008)}]{CavaliereTaylor08}
Cavaliere G, Taylor AMR (2008).
\newblock \enquote{Bootstrap Unit Root Tests for Time Series With Nonstationary
  Volatility.}
\newblock \emph{Econometric Theory}, \textbf{24}(1), 43--71.

\bibitem[{Cavaliere and Taylor(2009{\natexlab{a}})}]{CavaliereTaylor09ER}
Cavaliere G, Taylor AMR (2009{\natexlab{a}}).
\newblock \enquote{Bootstrap $M$ Unit Root Tests.}
\newblock \emph{Econometric Reviews}, \textbf{28}(5), 393--421.

\bibitem[{Cavaliere and
  Taylor(2009{\natexlab{b}})}]{cavaliere2009heteroskedastic}
Cavaliere G, Taylor AR (2009{\natexlab{b}}).
\newblock \enquote{Heteroskedastic Time Series With a Unit Root.}
\newblock \emph{Econometric Theory}, pp. 1228--1276.

\bibitem[{Chang(2002)}]{chang2002nonlinear}
Chang Y (2002).
\newblock \enquote{Nonlinear IV Unit Root Tests in Panels With Cross-Sectional
  Dependency.}
\newblock \emph{Journal of Econometrics}, \textbf{110}(2), 261--292.

\bibitem[{Chang and Park(2002)}]{ChangPark02}
Chang Y, Park JY (2002).
\newblock \enquote{On the Asymptotics of {ADF} Tests for Unit Roots.}
\newblock \emph{Econometric Reviews}, \textbf{21}, 431--447.

\bibitem[{Chang and Park(2003)}]{chang2003sieve}
Chang Y, Park JY (2003).
\newblock \enquote{A Sieve Bootstrap for the Test of a Unit Root.}
\newblock \emph{Journal of Time Series Analysis}, \textbf{24}(4), 379--400.

\bibitem[{Choi(2001)}]{choi2001unit}
Choi I (2001).
\newblock \enquote{Unit Root Tests for Panel Data.}
\newblock \emph{Journal of International Money and Finance}, \textbf{20}(2),
  249--272.

\bibitem[{Choi(2015)}]{Choi15}
Choi I (2015).
\newblock \emph{Almost All About Unit Roots: Foundations, Developments, and
  Applications}.
\newblock Cambridge University Press.

\bibitem[{Croissant and Millo(2008)}]{plmJSS}
Croissant Y, Millo G (2008).
\newblock \enquote{Panel Data Econometrics in \proglang{R}: The \pkg{plm}
  Package.}
\newblock \emph{Journal of Statistical Software}, \textbf{27}(2), 1--43.
\newblock \doi{10.18637/jss.v027.i02}.

\bibitem[{Croissant and Millo(2018)}]{plmbook}
Croissant Y, Millo G (2018).
\newblock \emph{Panel Data Econometrics with \proglang{R}}.
\newblock Wiley.

\bibitem[{Davidson and Flachaire(2008)}]{DavidsonFlachaire08}
Davidson R, Flachaire E (2008).
\newblock \enquote{The Wild Bootstrap, Tamed At Last.}
\newblock \emph{Journal of Econometrics}, \textbf{146}, 162--169.

\bibitem[{Dickey and Fuller(1979)}]{DickeyFuller79}
Dickey DA, Fuller WA (1979).
\newblock \enquote{Distribution of Estimators for Autoregressive Time Series
  With a Unit Root.}
\newblock \emph{Journal of the American Statistical Association},
  \textbf{74}(366a), 427--431.

\bibitem[{Dickey and Fuller(1981)}]{dickey1981likelihood}
Dickey DA, Fuller WA (1981).
\newblock \enquote{Likelihood Ratio Statistics for Autoregressive Time Series
  With a Unit Root.}
\newblock \emph{Econometrica}, pp. 1057--1072.

\bibitem[{Eddelbuettel(2013)}]{Rcpp13}
Eddelbuettel D (2013).
\newblock \emph{Seamless \proglang{R} and \proglang{C++} Integration with
  \pkg{Rcpp}}.
\newblock Springer-Verlag, New York.
\newblock \doi{10.1007/978-1-4614-6868-4}.
\newblock ISBN 978-1-4614-6867-7.

\bibitem[{Eddelbuettel and Balamuta(2017)}]{Rcpp17}
Eddelbuettel D, Balamuta JJ (2017).
\newblock \enquote{Extending \proglang{R} With \proglang{C++}: A Brief
  Introduction to \pkg{Rcpp}.}
\newblock \emph{PeerJ Preprints}, \textbf{5}, e3188v1.
\newblock ISSN 2167-9843.
\newblock \doi{10.7287/peerj.preprints.3188v1}.

\bibitem[{Eddelbuettel and Fran{\c{c}}ois(2011)}]{eddelbuettel2011rcpp}
Eddelbuettel D, Fran{\c{c}}ois R (2011).
\newblock \enquote{\pkg{Rcpp}: Seamless \proglang{R} and \proglang{C++}
  Integration.}
\newblock \emph{Journal of Statistical Software}, \textbf{40}(8), 1--18.

\bibitem[{Eddelbuettel and Sanderson(2014)}]{eddelbuettel2014rcpparmadillo}
Eddelbuettel D, Sanderson C (2014).
\newblock \enquote{\pkg{RcppArmadillo}: Accelerating \proglang{R} With
  High-Performance \proglang{C++} Linear Algebra.}
\newblock \emph{Computational Statistics \& Data Analysis}, \textbf{71},
  1054--1063.

\bibitem[{Elliott \emph{et~al.}(1996)Elliott, Rothenberg, and Stock}]{ERS96}
Elliott G, Rothenberg TJ, Stock JH (1996).
\newblock \enquote{Efficient Tests for an Autoregressive Unit Root.}
\newblock \emph{Econometrica}, \textbf{64}(4), 813--836.

\bibitem[{Enders(2008)}]{Enders08}
Enders W (2008).
\newblock \emph{Applied Econometric Time Series}.
\newblock 4th edition. John Wiley \& Sons.

\bibitem[{Friedrich \emph{et~al.}(2020)Friedrich, Smeekes, and Urbain}]{FSU20}
Friedrich M, Smeekes S, Urbain JP (2020).
\newblock \enquote{Autoregressive Wild Bootstrap Inference for Nonparametric
  Trends.}
\newblock \emph{Journal of Econometrics}, \textbf{214}(1), 81--109.

\bibitem[{Hansen(1995)}]{hansen1995rethinking}
Hansen BE (1995).
\newblock \enquote{Rethinking the Univariate Approach to Unit Root Testing:
  Using Covariates to Increase Power.}
\newblock \emph{Econometric Theory}, pp. 1148--1171.

\bibitem[{Harvey \emph{et~al.}(2009)Harvey, Leybourne, and Taylor}]{HLT09}
Harvey DI, Leybourne SJ, Taylor AMR (2009).
\newblock \enquote{Unit Root Testing in Practice: Dealing With Uncertainty Over
  the Trend and Initial Condition.}
\newblock \emph{Econometric Theory}, \textbf{25}(3), 587--636.

\bibitem[{Harvey \emph{et~al.}(2012)Harvey, Leybourne, and Taylor}]{HLT12}
Harvey DI, Leybourne SJ, Taylor AMR (2012).
\newblock \enquote{Testing for Unit Roots in the Presence of Uncertainty Over
  Both the Trend and Initial Condition.}
\newblock \emph{Journal of Econometrics}, \textbf{169}(2), 188--195.

\bibitem[{Holm(1979)}]{holm1979simple}
Holm S (1979).
\newblock \enquote{A Simple Sequentially Rejective Multiple Test Procedure.}
\newblock \emph{Scandinavian Journal of Statistics}, \textbf{6}, 65--70.

\bibitem[{Hylleberg \emph{et~al.}(1990)Hylleberg, Engle, Granger, and
  Yoo}]{hylleberg1990seasonal}
Hylleberg S, Engle RF, Granger CW, Yoo BS (1990).
\newblock \enquote{Seasonal Integration and Cointegration.}
\newblock \emph{Journal of Econometrics}, \textbf{44}(1-2), 215--238.

\bibitem[{Im \emph{et~al.}(2003)Im, Pesaran, and Shin}]{im2003testing}
Im KS, Pesaran MH, Shin Y (2003).
\newblock \enquote{Testing for Unit Roots in Heterogeneous Panels.}
\newblock \emph{Journal of Econometrics}, \textbf{115}(1), 53--74.

\bibitem[{Kleiber and Lupi(2011)}]{kleiber2011panel}
Kleiber C, Lupi C (2011).
\newblock \enquote{Panel Unit Root Testing With \proglang{R}.}
\newblock
  \urlprefix\url{https://r-forge.r-project.org/scm/viewvc.php/*checkout*/pkg/inst/doc/panelUnitRootWithR.pdf}.

\bibitem[{Kleiber and Lupi(2012)}]{punitroots}
Kleiber C, Lupi C (2012).
\newblock \emph{\pkg{punitroots}: Tests for Unit Roots in Panels of (Economic)
  Time Series, With and Without Cross-sectional Dependence}.
\newblock \proglang{R} package version 0.0-2,
  \urlprefix\url{https://r-forge.r-project.org/projects/punitroots/}.

\bibitem[{Kreiss \emph{et~al.}(2011)Kreiss, Paparoditis, and Politis}]{KPP11}
Kreiss JP, Paparoditis E, Politis DN (2011).
\newblock \enquote{On the Range of Validity of the Autoregressive Sieve
  Bootstrap.}
\newblock \emph{The Annals of Statistics}, \textbf{39}, 2103--2130.

\bibitem[{Kwiatkowski \emph{et~al.}(1992)Kwiatkowski, Phillips, Schmidt, and
  Shin}]{kwiatkowski1992testing}
Kwiatkowski D, Phillips PC, Schmidt P, Shin Y (1992).
\newblock \enquote{Testing the Null Hypothesis of Stationarity Against the
  Alternative of a Unit Root.}
\newblock \emph{Journal of Econometrics}, \textbf{54}(1-3), 159--178.

\bibitem[{Levin \emph{et~al.}(2002)Levin, Lin, and Chu}]{levin2002unit}
Levin A, Lin CF, Chu CSJ (2002).
\newblock \enquote{Unit Root Tests in Panel Data: Asymptotic and Finite-Sample
  Properties.}
\newblock \emph{Journal of Econometrics}, \textbf{108}(1), 1--24.

\bibitem[{L\'{o}pez-de Lacalle and Boshnakov(2020)}]{uroot}
L\'{o}pez-de Lacalle J, Boshnakov GN (2020).
\newblock \emph{\pkg{uroot}: Unit Root Tests for Seasonal Time Series}.
\newblock \proglang{R} package version 2.1-2,
  \urlprefix\url{https://CRAN.R-project.org/package=uroot}.

\bibitem[{Lupi(2009)}]{lupi2009unit}
Lupi C (2009).
\newblock \enquote{Unit Root CADF Testing With \proglang{R}.}
\newblock \emph{Journal of Statistical Software}, \textbf{32}(2), 1--19.

\bibitem[{Maddala and Wu(1999)}]{maddala1999comparative}
Maddala GS, Wu S (1999).
\newblock \enquote{A Comparative Study of Unit Root Tests With Panel Data and a
  New Simple Test.}
\newblock \emph{Oxford Bulletin of Economics and Statistics}, \textbf{61}(S1),
  631--652.

\bibitem[{Mallet(2017)}]{URT}
Mallet O (2017).
\newblock \emph{\pkg{URT}: Fast Unit Root Tests and OLS Regression in
  \proglang{C++} With Wrappers for \proglang{R} and \proglang{Python}}.
\newblock \urlprefix\url{https://github.com/olmallet81/URT}.

\bibitem[{Mammen(1993)}]{Mammen93}
Mammen E (1993).
\newblock \enquote{Bootstrap and Wild Bootstrap for High Dimensional Linear
  Models.}
\newblock \emph{The Annals of Statistics}, \textbf{21}, 255--285.

\bibitem[{McCracken and Ng(2020)}]{mccracken2020fred}
McCracken M, Ng S (2020).
\newblock \enquote{FRED-QD: A Quarterly Database for Macroeconomic Research.}
\newblock \emph{Working Paper 26872}, National Bureau of Economic Research.

\bibitem[{McCracken and Ng(2016)}]{McCrackenNg16}
McCracken MW, Ng S (2016).
\newblock \enquote{FRED-MD: A Monthly Database for Macroeconomic Research.}
\newblock \emph{Journal of Business \& Economic Statistics}, \textbf{34}(4),
  574--589.

\bibitem[{Moon and Perron(2012)}]{MoonPerron12}
Moon HR, Perron B (2012).
\newblock \enquote{Beyond Panel Unit Root Tests: Using Multiple Testing to
  Determine the Non stationarity Properties of Individual Series in a Panel.}
\newblock \emph{Journal of Econometrics}, \textbf{169}(1), 29--33.

\bibitem[{M{\"u}ller and Elliott(2003)}]{MuellerElliott03}
M{\"u}ller UK, Elliott G (2003).
\newblock \enquote{Tests for Unit Roots and the Initial Condition.}
\newblock \emph{Econometrica}, \textbf{71}(4), 1269--1286.

\bibitem[{Ng and Perron(2001)}]{NgPerron01}
Ng S, Perron P (2001).
\newblock \enquote{Lag Length Selection and the Construction of Unit Root Tests
  With Good Size and Power.}
\newblock \emph{Econometrica}, \textbf{69}(6), 1519--1554.

\bibitem[{Ng and Perron(2005)}]{NgPerron05}
Ng S, Perron P (2005).
\newblock \enquote{A Note on the Selection of Time Series Models.}
\newblock \emph{Oxford Bulletin of Economics and Statistics}, \textbf{67},
  115--134.

\bibitem[{Palm \emph{et~al.}(2008)Palm, Smeekes, and Urbain}]{PSU08}
Palm FC, Smeekes S, Urbain JP (2008).
\newblock \enquote{Bootstrap Unit Root Tests: Comparison and Extensions.}
\newblock \emph{Journal of Time Series Analysis}, \textbf{29}(1), 371--401.

\bibitem[{Palm \emph{et~al.}(2011)Palm, Smeekes, and Urbain}]{PSU11}
Palm FC, Smeekes S, Urbain JP (2011).
\newblock \enquote{Cross-Sectional Dependence Robust Block Bootstrap Panel Unit
  Root Tests.}
\newblock \emph{Journal of Econometrics}, \textbf{163}(1), 85--104.

\bibitem[{Pantula(1989)}]{Pantula89}
Pantula SG (1989).
\newblock \enquote{Testing for Unit Roots in Time Series Data.}
\newblock \emph{Econometric Theory}, \textbf{5}(2), 256--271.

\bibitem[{Paparoditis and Politis(2003)}]{paparoditis2003residual}
Paparoditis E, Politis DN (2003).
\newblock \enquote{Residual-Based Block Bootstrap for Unit Root Testing.}
\newblock \emph{Econometrica}, \textbf{71}(3), 813--855.

\bibitem[{Paparoditis and Politis(2005)}]{PapPol05}
Paparoditis E, Politis DN (2005).
\newblock \enquote{Bootstrapping Unit Root Tests for Autoregressive Time
  Series.}
\newblock \emph{Journal of the American Statistical Association}, \textbf{100},
  545--553.

\bibitem[{Paparoditis and Politis(2018)}]{paparoditis2018the}
Paparoditis E, Politis DN (2018).
\newblock \enquote{The Asymptotic Size and Power of the Augmented
  {D}ickey-{F}uller Test for a Unit Root.}
\newblock \emph{Econometric Reviews}, \textbf{37}(9), 955--973.

\bibitem[{Perron and Qu(2007)}]{perron2007simple}
Perron P, Qu Z (2007).
\newblock \enquote{A Simple Modification to Improve the Finite Sample
  Properties of Ng and Perron's Unit Root Tests.}
\newblock \emph{Economics Letters}, \textbf{94}(1), 12--19.

\bibitem[{Pesaran(2007)}]{pesaran2007simple}
Pesaran MH (2007).
\newblock \enquote{A simple panel unit root test in the presence of
  cross-section dependence.}
\newblock \emph{Journal of Applied Econometrics}, \textbf{22}(2), 265--312.

\bibitem[{Pesaran(2012)}]{Pesaran12}
Pesaran MH (2012).
\newblock \enquote{On the Interpretation of Panel Unit Root Tests.}
\newblock \emph{Economics Letters}, \textbf{116}(3), 545--546.

\bibitem[{Pfaff(2008)}]{urcabook}
Pfaff B (2008).
\newblock \emph{Analysis of Integrated and Cointegrated Time Series with
  \proglang{R}}.
\newblock Second edition. Springer-Verlag, New York.
\newblock ISBN 0-387-27960-1, \urlprefix\url{http://www.pfaffikus.de}.

\bibitem[{Phillips and Perron(1988)}]{phillips1988testing}
Phillips PC, Perron P (1988).
\newblock \enquote{Testing for a Unit Root in Time Series Regression.}
\newblock \emph{Biometrika}, \textbf{75}(2), 335--346.

\bibitem[{\proglang{R} Core~Team(2021)}]{R}
\proglang{R} Core~Team (2021).
\newblock \emph{\proglang{R}: {A} Language and Environment for Statistical
  Computing}.
\newblock \proglang{R} Foundation for Statistical Computing, Vienna, Austria.
\newblock \urlprefix\url{https://www.R-project.org/}.

\bibitem[{Psaradakis(2001)}]{Psaradakis01}
Psaradakis Z (2001).
\newblock \enquote{Bootstrap Tests for an Autoregressive Unit Root in the
  Presence of Weakly Dependent Errors.}
\newblock \emph{Journal of Time Series Analysis}, \textbf{22}, 577--594.

\bibitem[{Rho and Shao(2019)}]{RhoShao19}
Rho Y, Shao X (2019).
\newblock \enquote{Bootstrap-Assisted Unit Root Testing With Piecewise Locally
  Stationary Errors.}
\newblock \emph{Econometric Theory}, \textbf{35}(1), 142--166.

\bibitem[{Romano \emph{et~al.}(2008{\natexlab{a}})Romano, Shaikh, and
  Wolf}]{RSW08fdr}
Romano JP, Shaikh AM, Wolf M (2008{\natexlab{a}}).
\newblock \enquote{Control of the False Discovery Rate Under Dependence Using
  the Bootstrap and Subsampling.}
\newblock \emph{Test}, \textbf{17}(3), 417--442.

\bibitem[{Romano \emph{et~al.}(2008{\natexlab{b}})Romano, Shaikh, and
  Wolf}]{RSW08}
Romano JP, Shaikh AM, Wolf M (2008{\natexlab{b}}).
\newblock \enquote{Formalized Data Snooping Based on Generalized Error Rates.}
\newblock \emph{Econometric Theory}, \textbf{24}(2), 404--447.

\bibitem[{Romano and Wolf(2005)}]{RomanoWolf05}
Romano JP, Wolf M (2005).
\newblock \enquote{Stepwise Multiple Testing As Formalized Data Snooping.}
\newblock \emph{Econometrica}, \textbf{73}(4), 1237--1282.

\bibitem[{Schwert(1989)}]{Schwert89}
Schwert GW (1989).
\newblock \enquote{Tests for Unit Roots: A Monte Carlo Investigation.}
\newblock \emph{Journal of Business and Economic Statistics}, \textbf{7}(1),
  147--159.

\bibitem[{Shao(2010)}]{Shao10}
Shao X (2010).
\newblock \enquote{The Dependent Wild Bootstrap.}
\newblock \emph{Journal of the American Statistical Association},
  \textbf{105}(489), 218--235.

\bibitem[{Shao(2011)}]{shao2011bootstrap}
Shao X (2011).
\newblock \enquote{A Bootstrap-Assisted Spectral Test of White Noise Under
  Unknown Dependence.}
\newblock \emph{Journal of Econometrics}, \textbf{162}(2), 213--224.

\bibitem[{Smeekes(2013)}]{smeekes2013detrending}
Smeekes S (2013).
\newblock \enquote{Detrending Bootstrap Unit Root Tests.}
\newblock \emph{Econometric Reviews}, \textbf{32}(8), 869--891.

\bibitem[{Smeekes(2015)}]{Smeekes15}
Smeekes S (2015).
\newblock \enquote{Bootstrap Sequential Tests to Determine the Order of
  Integration of Individual Units in a Time Series Panel.}
\newblock \emph{Journal of Time Series Analysis}, \textbf{36}(3), 398--415.

\bibitem[{Smeekes and Taylor(2012)}]{SmeekesTaylor12}
Smeekes S, Taylor AMR (2012).
\newblock \enquote{Bootstrap Union Tests for Unit Roots in the Presence of
  Nonstationary Volatility.}
\newblock \emph{Econometric Theory}, \textbf{28}(2), 422--456.

\bibitem[{Smeekes and Urbain(2014{\natexlab{a}})}]{SmeekesUrbain14RM}
Smeekes S, Urbain JP (2014{\natexlab{a}}).
\newblock \enquote{A Multivariate Invariance Principle for Modified Wild
  Bootstrap Methods With an Application to Unit Root Testing.}
\newblock \emph{GSBE Research Memorandum RM/14/008}, Maastricht University.

\bibitem[{Smeekes and Urbain(2014{\natexlab{b}})}]{SmeekesUrbain14}
Smeekes S, Urbain JP (2014{\natexlab{b}}).
\newblock \enquote{On the Applicability of the Sieve Bootstrap in Time Series
  Panels.}
\newblock \emph{Oxford Bulletin of Economics and Statistics}, \textbf{76}(1),
  139--151.

\bibitem[{Smeekes and Wijler(2020)}]{SmeekesWijler20}
Smeekes S, Wijler E (2020).
\newblock \enquote{Unit Roots and Cointegration.}
\newblock In P~Fuleky (ed.), \emph{Macroeconomic Forecasting in the Era of Big
  Data}, volume~52 of \emph{Advanced Studies in Theoretical and Applied
  Econometrics}, chapter~17, pp. --541--584. Springer-Verlag.

\bibitem[{Smeekes and Wilms(2021)}]{bootUR}
Smeekes S, Wilms I (2021).
\newblock \emph{\pkg{bootUR}: Bootstrap Unit Root Tests}.
\newblock \proglang{R} package version 0.4.2,
  \urlprefix\url{https://CRAN.R-project.org/package=bootUR}.

\bibitem[{Theu{\ss}l and Zeileis(2009)}]{theussl2009collaborative}
Theu{\ss}l S, Zeileis A (2009).
\newblock \enquote{Collaborative Software Development Using
  \proglang{R-Forge}.}
\newblock \emph{The \proglang{R} Journal}, \textbf{1}(1), 9--14.

\bibitem[{Trapletti \emph{et~al.}(2021)Trapletti, Hornik, and
  LeBaron}]{tseries}
Trapletti A, Hornik K, LeBaron B (2021).
\newblock \emph{\pkg{tseries}: Time Series Analysis and Computational Finance}.
\newblock \proglang{R} package version 0.10-49,
  \urlprefix\url{https://CRAN.R-project.org/package=tseries}.

\bibitem[{Tsung-wu(2019)}]{pdR}
Tsung-wu H (2019).
\newblock \emph{\pkg{pdR}: Threshold Model and Unit Root Tests in Cross-Section
  and Time Series Data}.
\newblock \proglang{R} package version 1.7,
  \urlprefix\url{https://CRAN.R-project.org/package=pdR}.

\bibitem[{Wuertz \emph{et~al.}(2017)Wuertz, Setz, and Chalabi}]{fUnitRoots}
Wuertz D, Setz T, Chalabi Y (2017).
\newblock \emph{\pkg{fUnitRoots}: \pkg{Rmetrics} - Modelling Trends and Unit
  Roots}.
\newblock \proglang{R} package version 3042.79,
  \urlprefix\url{https://CRAN.R-project.org/package=fUnitRoots}.

\bibitem[{Zhang and Cheng(2014)}]{zhang2014bootstrapping}
Zhang X, Cheng G (2014).
\newblock \enquote{Bootstrapping High Dimensional Time Series.}
\newblock ArXiv e-print 1406.1037.

\bibitem[{Zhang \emph{et~al.}(2011)Zhang, Yu, and McLeod}]{mleurwp}
Zhang Y, Yu H, McLeod AI (2011).
\newblock \enquote{Maximum Likelihood Unit Root Test.}
\newblock \emph{Working Paper}.

\end{thebibliography}

\appendix

\section{Simulation Study} \label{appendix:simstudy}
We perform a simulation study to investigate the small sample performance of the different unit root tests, thereby largely following the design choices in \cite{Smeekes15}.
We consider three designs: a
(i) micro panel with $T=25$ and $N=200$
(ii) macro panel with $T=100$ and $N=50$, and a
(iii) high-dimensional panel with $T=100$ and $N=100$.
Data are generated from model  \eqref{dgp} with $\beta = 0$ (hence no deterministic components), and the error terms drawn from a non-central $t$~distribution with non-centrality parameter 10 and 4 degrees of freedom. For the stationary units $i\leq k_0$ where $k_0 = \lfloor q_0 N \rfloor$ in the panel, we set the AR  parameter $\rho = 0.6$ in setting (i) and $\rho = 0.9$ in settings (ii) and (iii). 
For the non-stationary units $i > k_0$, we take $\rho = 1$.
As the true proportions of stationary units, we consider $q_0 = 0, 0.2, 0.5, 0.9$.

Results are reported for the asymptotic \fct{adf} test, and the bootstrap \fct{boot\_ur}, \fct{boot\_sqt} and \fct{boot\_panel} tests. For all tests,  we apply the \fct{order\_integration} function with \code{max\_order = 1} to test for a unit root in the series in levels. For the bootstrap tests, we use \code{B = 499} bootstrap replications. For the \fct{boot\_sqt} test, we take \code{steps = 0:8/8} in the micro and high-dimensional panel, \code{steps = 0:4/4} for the macro panel. All other arguments are kept at their defaults. Results are based on 200 Monte Carlo simulations.

We assess the performance of the unit root tests on three metrics.
We report the average proportion of units 
(1) correctly classified (`correct'),
(2) incorrectly found to be stationary (`size'),
(3) correctly found to be stationary (`power').
Results are reported in Table \ref{table:sim-study}. 

\begin{table}[t]
\resizebox{0.76\textwidth}{!}{\begin{minipage}{\textwidth}
\begin{tabular}{lllcccccccccccccc}
  \hline
design & unit root && \multicolumn{2}{c}{\underline{$q_0 = 0$}} && \multicolumn{3}{c}{\underline{$q_0 = 0.2$}} && \multicolumn{3}{c}{\underline{$q_0 = 0.5$}} && \multicolumn{3}{c}{\underline{$q_0 = 0.9$}} \\
 &test&& correct & size  && correct & size & power && correct & size & power && correct & size & power \\ 
  \hline
micro & \code{adf} && 0.96 & 0.04 && 0.82 & 0.04 & 0.26 && 0.61 & 0.04 & 0.26 && 0.33 & 0.04 & 0.26 \\ 
& \code{boot\_ur} && 0.94 & 0.06 && 0.84 & 0.06 & 0.44 && 0.69 & 0.06 & 0.45 && 0.49 & 0.06 & 0.44 \\ 
& \code{boot\_sqt} && 1.00 & 0.00 && 0.83 & 0.04 & 0.31 && 0.68 & 0.06 & 0.42 && 0.57 & 0.09 & 0.53 \\ 
& \code{boot\_fdr} && 1.00 & 0.00 && 0.80 & 0.00 & 0.00 && 0.50 & 0.00 & 0.00 && 0.11 & 0.00 & 0.01 \\ 
   \hline
macro & \code{adf} && 0.96 & 0.04 && 0.81 & 0.04 & 0.23 && 0.60 & 0.04 & 0.23 && 0.30 & 0.04 & 0.23 \\ 
& \code{boot\_ur} && 0.94 & 0.06 && 0.82 & 0.06 & 0.36 && 0.65 & 0.06 & 0.35 && 0.42 & 0.07 & 0.36 \\ 
& \code{boot\_sqt} && 0.99 & 0.01 && 0.80 & 0.08 & 0.31 && 0.68 & 0.12 & 0.49 && 0.70 & 0.28 & 0.69 \\ 
& \code{boot\_fdr} && 1.00 & 0.00 && 0.80 & 0.00 & 0.03 && 0.52 & 0.00 & 0.03 && 0.15 & 0.01 & 0.06 \\ 
   \hline 
high-dim & \code{adf} && 0.96 & 0.04 && 0.81 & 0.04 & 0.23 && 0.60 & 0.04 & 0.24 && 0.30 & 0.04 & 0.23 \\ 
&  \code{boot\_ur} && 0.94 & 0.06 && 0.82 & 0.06 & 0.36 && 0.65 & 0.06 & 0.36 && 0.42 & 0.06 & 0.36 \\ 
&  \code{boot\_sqt} && 0.99 & 0.01 && 0.81 & 0.05 & 0.27 && 0.65 & 0.06 & 0.37 && 0.51 & 0.10 & 0.46 \\ 
& \code{boot\_fdr} && 1.00 & 0.00 && 0.80 & 0.00 & 0.01 && 0.51 & 0.00 & 0.02 && 0.13 & 0.00 & 0.03 \\ 
   \hline   
\end{tabular}
\end{minipage} }
\caption{Simulation results for the four unit root tests across the different simulation designs (micro, macro and high-dim panel), as well as varying proportion of stationary units $q_0$.} \label{table:sim-study}
\end{table}

For the micro panel, all tests produce a roughly similar, high proportion of correct classifications and a fairly low probability of classifying units incorrectly as $I(0)$ when there are few stationary units. All tests have rather low power, which is expected given the small number of time series observations, but this is most severe for the \fct{boot\_fdr} test.  As the proportion of stationary units increases, \code{boot\_sqt} starts to perform better than the other tests; the opposite occurs for \code{boot\_fdr}.

For the macro and high-dimensional panel, results for the tests \code{adf}, \code{boot\_ur} and \code{boot\_fdr} are largely comparable to those of the micro panel. 
The \code{boot\_sqt} test selects too many stationary units for the macro panel (especially when their true proportion is large), which occurs due to the choice of the argument \code{steps} set to four equally spaced quantiles. Note that the \code{boot\_fdr} test still displays very low power in the macro panel, whereas its power reported in \cite{Smeekes15} is considerably better.
Note that we consider a different value under the alternative here; we use a fixed AR parameter of 0.9 whereas \citealp{Smeekes15} drew the AR parameter from a uniform distribution between zero and 0.9. In addition, the usage of the union of rejection test instead of the ADF test with intercept in \citealp{Smeekes15} as well as the different choice of the maximum number of lagged differences in the ADF regression (we use the sample size dependent default in \pkg{bootUR}, 12 in this case, whereas the maximum was set to four in \citealp{Smeekes15}) may also affect the performance of the \code{boot\_fdr} test.
\end{document}